\documentclass{JHEP}
\usepackage{epsfig}
\usepackage{amssymb,latexsym,amsfonts}

\newcommand{\be}{\begin{equation}}
\newcommand{\ee}{\end{equation}}
\newcommand{\bd}{\begin{displaymath}}
\newcommand{\ed}{\end{displaymath}}
\newcommand{\ba}{\begin{array}}
\newcommand{\ea}{\end{array}}
\newcommand{\bt}{\begin{tabular}}
\newcommand{\et}{\end{tabular}}

\newcommand{\bea}{\begin{eqnarray}}
\newcommand{\eea}{\end{eqnarray}}

\newcommand{\Z}{\mathbb{Z}}

\newcommand{\R}{\mathbb{R}}

\newcommand{\p}{\mathbb{P}}

\newcommand{\mod}{\textrm{mod}}

\newcommand{\e}{\epsilon}

\newcommand{\dif}{\textrm{d}}
\preprint{LPT-ENS 01/32 \\ PAR-LPTHE 01-31\\ {\tt
hep-th/0106267}}

\vspace{.5cm}

\title{Classifying orientifolds by flat $n$-gerbes}

\author{Arjan Keurentjes
\\ {\it LPTHE, Universit\'e Pierre et Marie Curie, Paris VI\\
4 place Jussieu, Tour 16, F-75252, Paris Cedex 05, France}\\
\vspace{0.3cm}{\it Laboratoire de Physique Th\'eorique de l'Ecole
Normale Sup\'erieure\\ 24 rue Lhomond, F-75231 Paris Cedex 05,
France}\\ \email{arjan@lpthe.jussieu.fr} }

\abstract{The discrete tensorial charges carried by orientifold planes 
define $n$-gerbes in space-time. The simplest way to ensure a consistent 
string compactification is to require these gerbes to be flat. This 
results in expressions for the local gerbe-holonomies around each 
orientifold plane, describing its charges. Inverting the procedure and 
considering all flat gerbes leads to a classification of orientifold 
configurations. Requiring that the tadpole is cancelled by adding 
D-branes, we classify all supersymmetric orientifolds on $T^k/\Z_2$ 
with $2^k$ O$(9-k)$ planes at the fixed points, for $k \leq 6$. For $k=6$ 
these theories organize in orbits of the $SL(2,\Z)$ S-duality symmetry of 
${\cal N}=4$ supersymmetric gauge theories.} 

\begin{document}
\section{Introduction}

An orientifold is a construction in string theory, which involves the 
gauging of a symmetry of the theory which is a combination of a symmetry 
of the target space together with the orientation reversal of the string 
world sheet \cite{Sagnotti, Horava}. If the target space symmetry has a fixed 
point set, its connected components are called orientifold fixed planes, or 
just orientifold planes for short (also orientifold lines and points, for 
sufficiently low dimension of the orientifold plane). 

In this article we will restrict slightly further to the best studied 
category of orientifold planes, the ones that appear in type II string 
theories and locally look like $\R^{p,1} \times (\R^{9-p})/\Z_2$, where 
the $\Z_2$ inverts the coordinates on $\R^{9-p}$ (we will review the 
full action in the next section). These planes preserve 16 of the 32 
supersymmetries of the type II theories.

Orientifold planes have various interesting properties, that are 
intrinsically related. A stack of $n$ D-branes has generically a $U(n)$ 
gauge theory propagating on its worldvolume; when the stack of branes is 
coinciding with an orientifold plane, the group is projected to $O(n)$ 
or $Sp(n/2)$ depending on the kind of orientifold plane. Whether the 
group is projected to $O(n)$ or $Sp(n/2)$ is correlated with a discrete 
charge carried by the orientifold plane. Another charge distinguishes 
between $O(2n)$ and $O(2n+1)$ symmetry. Although in principle more 
charges are possible \cite{Hanany00}, it is not clear what the effect of 
such charges is on string perturbation theory (if such an effect can be 
treated in perturbation theory at all), and we will 
restrict throughout this paper to the two charges mentioned. 

The charges are tensorial in nature; they result in fluxes of certain 
tensorfields. The spectrum of charges can be studied in a suitable 
formalism such as cohomology or K-theory \cite{Bergman}. In this note we 
wish to study another aspect of the presence of such charges. In the 
same sense as that electric and magnetic charges give rise to 
electromagnetic fields, which can in turn be thought of in terms of a 
connection on and curvature of bundles, the tensorial charges carried by 
orientifold planes define tensorfields over space-time. As, in form 
notation, the one form and higher form tensorfields are very similar, it 
is tempting to also give a description of higher form tensorfields as 
connections on ``something''. A particular formalism that is suited for 
this is the idea of (Abelian) gerbes \cite{Hitchin} (see also \cite{Sharpe} 
for a discussion with applications closer to the one in this paper). 
Gerbes are a generalization of bundles. As a matter of fact, there is a whole 
tower of mathematical objects, called $n$-gerbes.  

The discrete tensorial charges of the orientifold planes 
give rise to $n$-gerbes. For a consistent string background, we have to 
satisfy the equations of motion. The easiest way to do so is to 
require the gerbes to be flat, and this is the problem we will study in 
this paper. A single orientifold plane does not give us any problems, 
its discrete charges are derived from a cohomology analysis that 
implicitly tells us that the curvature of the gerbe is at most torsion. 
For a composition of multiple orientifold planes in a certain geometry 
this is no longer necessarily true. The analysis in terms of a flux 
determined by the cohomology of the surrounding space at infinity is no 
longer valid; the discrete identifications that come with 
multiple orientifold planes modify the topology of the space at 
infinity. 

In this paper we will study configurations of orientifold planes on the 
space $\R^{p,1} \times (T^{9-p})/\Z_2$. The orientifold planes at the fixed 
points are allowed to carry two kinds of discrete charges. 
We will review some properties of the relevant orientifold planes in section 
\ref{Oplanes}. Using some technology developed in a previous paper 
\cite{deBoer}, it is actually not hard to write down the possible flat gerbes. 
We will do so in section \ref{ghol}. 

The constraints we will obtain are not very restrictive for large 
$p$, but will become increasingly so for smaller values of $p$. 
We conjecture that, at least for the configurations we are 
studying, there is only one more criterion needed for a consistent 
compactification. This is the requirement that the RR $(p+1)$-- form tadpole 
be cancelled. This can always be done by adding a suitable number of 
D$p$- or anti D$p$-branes. The most non-suspect backgrounds are the 
supersymmetric ones, and these require the addition of D$p$-branes. We 
will restrict mostly to these supersymmetric orientifolds, although we 
will occasionally comment on their non-supersymmetric cousins.

With the restrictions made, we first study orientifolds with two 
kinds of planes. and classify these up to geometrical symmetries. This 
will be done in section \ref{twoplanes}. The more elaborate case where 
we no longer restrict to a subset of the possible kinds of planes will 
be taken up in section \ref{BC}. The results of these sections, a 
complete classification for $p \geq 3$, can be found in our tables 
\ref{NScharges}, \ref{RRcharges}, \ref{mixeda} and \ref{mixedb}. Due to 
a technical problem we cannot proceed beyond 4 noncompact dimensions. In 
4 dimensions, the orientifolds we find lead to ${\cal N}=4$  
supersymmetric gauge theories. We compare our data with the requirement 
of the S-duality of these theories in section \ref{S}. Finally we 
summarize and conclude in section \ref{sum}.

\section{Aspects of orientifold planes} \label{Oplanes}

The orientifold planes we are interested in locally look like $\R^{p,1} 
\times (\R^{9-p})/\Z_2$. The generator of the relevant $\Z_2$ 
transformation is a product of three factors
\bd
I \cdot \Omega \cdot J
\ed
where $I$ acts as inversion on the $9-p$ spatial coordinates of 
$\R^{9-p}$ torus, and $\Omega$ is the worldsheet parity operator. 
Finally, $J$ is either the identity operator for $p=0,1 \  \mod \ 4$, or 
$(-)^{F_L}$ for $p=2,3 \  \mod \  4$, with $F_L$ the left moving fermion 
number. For $p$ even resp. odd, this $\Z_2$ is a symmetry of IIA resp. IIB 
theory. As a consequence, in these theories one can find orientifold 
planes of spatial dimension $p$, subsequently denoted as O$p$ planes.
   
For a single O$p$ plane, the space at infinity is $S^{8-p}/\Z_2 = \R 
\p^{8-p}$. Possible tensorial fluxes coming from the orientifold plane 
are traditionally classified by the cohomologies\footnote{According to 
current lore, one should use K-theory for the RR fluxes \cite{Bergman}. For 
the orientifold planes we will consider this gives the same result as 
cohomology.} of $\R \p^{8-p}$. We will be interested in orientifold planes 
with a flux from the NS-NS $B$-field. The operators $I$ and $J$ acts trivially 
on this form, but it receives a minus sign from the operator $\Omega$. It is 
therefore classified by \be [ \dif B ] = [ H ] \ \in \ H^3(\R \p^{8-p}, 
\tilde{\Z}) = \Z_2 \ee
This only makes sense for $p \leq 6$, but for higher values of $p$ one 
may work with equivariant cohomology \cite{deBoer, Witten97}, which 
also results in a $\Z_2$ spectrum. The two elements of $\Z_2$ correspond 
to different signs for a closed string diagram. As is well known, 
this is correlated with the action of the orientifold projection on the 
Chan-Paton matrices of the open string. The trivial element of $\Z_2$ 
results in an $O(n)$ gauge group\footnote{Actually, due to subtle 
effects that are non-perturbative from the viewpoint of the type II 
description, the actual gauge group is not (necessarily) $O(n)$. For 
O$9^-$, the actual gauge group is $Spin(32)/\Z_2$ \cite{Witten98b}. For lower 
values of $p$ in O$p^-$ there are in general multiple possibilities, among 
which $O(n)$, $SO(n)$, $Spin(n)$. The topology of this group does not even 
have to be such that it can be embedded in $Spin(32)/\Z_2$ \cite{deBoer}. We 
will simply write down the group that is manifest in perturbation theory, 
which is $O(n)$, and ignore symmetry breakings and/or extensions by 
non-perturbative effects.}  on D$p$-branes, and the non-trivial one in an 
$Sp(n/2)$ gauge group. In the first case, we will denote the plane as O$p^-$, 
in the second case as O$p^+$.

We will also distinguish an RR charge. Normally the effect of RR charges 
would be hard to describe in string perturbation theory. The 
exception to the rule was discovered first for O$3$ 
planes \cite{Witten98a}. The $SL(2,\Z)$ duality of type IIB theories 
translates to $S$-duality of the ${\cal N}=4$ supersymmetric Yang-Mills 
theory that is found on the D$3$ brane. In the presence of an O$3^+$ 
plane, this is an $Sp(k)$ gauge theory. S-duality maps this to an 
$SO(2k+1)$ gauge theory \cite{Goddard}, and the NS-NS 2--form charge to an RR 
2-form charge. By T-duality, this should extend to other O$p$ planes with $p 
\neq 3$.

Indeed, one can define a RR $(5-p)$--form charge for O$p$ planes. The 
relevant cohomologies  are given by
\bea
\left[ \dif C^{5-p} \right] = \left[ G^{6-p} \right]  & \in & H^{6-p}(\R 
\p^{8-p}, \Z)=\Z_2, \qquad p \textrm{ even}\\ 
\left[ \dif C^{5-p} \right] = \left[ G^{6-p} \right] & \in & 
H^{6-p}(\R \p^{8-p}, \tilde{\Z})=\Z_2. \qquad p \textrm{ odd}
\eea
This puts a natural upper bound at $p=5$. Above this bound, one still 
may use T-duality, but the result is that the RR flux is no longer 
localized at the plane.

There is a simple realization of a plane with such a flux: As $SO(2k+1)$ 
is the gauge group that is found on an O$p^-$ plane with an odd number 
of D$p$ branes on it, and as even number of branes can always be moved 
away from the plane, there is obviously a bound state of an O$p^-$ with 
a single D$p$-brane. We will reserve a special notation for this bound 
state, calling it $\widetilde{\textrm{O}p}{}^-$. The tilde is meant to 
indicate the non-trivial RR flux associated to the plane. The RR $(p+1)$--form 
charge of this plane is the sum of one unit of D$p$ brane charge and the 
charge of the O$p^-$ plane. This shift in $(p+1)$--form charge can also be 
understood from K-theory \cite{Bergman}. The upper bound of $p=5$ seems 
to suggest that $\widetilde{\textrm{O}p}{}^-$ planes are absent for $p > 5$. 
An attempt to evade this conclusion can be found in \cite{Hyakutake}. The 
approach of this paper requires the introduction of a discrete 
cosmological constant, which would break supersymmetry, and should 
modify the Einstein equation, excluding flat space (away from 
the orientifold plane) as a solution. It therefore falls outside the 
category of planes that we are interested in in this paper.

The final plane is the $\widetilde{\textrm{O}p}{}^+$. This carries both the 
NS-NS 2--form charge and the non-trivial RR $(5-p)$--form charge. It is not 
possible to form a bound state of an O$p^+$ with an odd number of 
D$p$-branes. Therefore there is no reason to expect a shift in the RR 
$(p+1)$--form charge, and also a K-theory computation does not indicate 
this \cite{Bergman}. It is actually not so easy to distinguish O$p^+$ and 
$\widetilde{\textrm{O}p}{}^+$. For $p=3$ some differences can be seen in the 
non-perturbative spectrum of monopoles and dyons in the ${\cal N}=4$ 
theories that arise when a stack of D3 branes coincides with the O$3$ 
plane \cite{Hanany00}. More on $\widetilde{\textrm{O}p}{}^+$ planes can be found 
in \cite{Gimon98, Hanany98, Hanany00, Hori, Keurentjes01, Witten98a}. 

The planes that are relevant to this paper, and their properties are 
summarized in the following table
\bd
\begin{tabular}{|c||c|c|c|c|}
\hline
O$p$ plane & group & $R_{p+1}$  & ${\cal B}$ & ${\cal C}$ \\
\hline
O$p^-$ & $O(2k)$ &  $-2^{p-4}$ & 0 & 0 \\
O$p^+$ & $Sp(k)$  & $+ 2^{p-4}$ & 1 & 0 \\
$\widetilde{\textrm{O}p}{}^-$ & $O(2k+1)$ & $1-2^{p-4}$ & 0 & 1\\
$\widetilde{\textrm{O}p}{}^+$ & $Sp(k)$ & $+ 2^{p-4}$ & 1 & 1 \\
\hline
\end{tabular}
\ed
The last three columns denote the charges of a plane: in the third 
column the RR $(p+1)$--form charge, denoted by $R_{p+1}$, in the fourth the 
discrete NS-NS 2--form charge, which we will denote by ${\cal B}$ in the 
remainder of this paper, while the last column denotes the discrete RR 
$(5-p)$--form charge ${\cal C}$. 

\section{Gerbe holonomies and orientifold charges} \label{ghol}

For a single O$p$ plane its charges can in principle be measured by 
surrounding it by a sufficiently large (hyper)surface. For example, it 
should be possible to measure ${\cal B}$ by computing $\int B_2$ over a 
suitably chosen 2-surface. For a single O$p$-plane this causes no 
problems. For a configuration of multiple planes, the various fluxes of 
the planes have to be patched together. To achieve this, we turn to the 
formalism of (Abelian) gerbes \cite{Hitchin}. Some of the material 
presented here appeared (in a slightly different form, and adapted to a 
different problem) in \cite{deBoer}. We will repeat some of the 
discussion because the formalism of gerbes has entered the physics 
literature rather recently, and to make the paper more self-contained.

Consider an $n$-gerbe\footnote{It is customary to call 1-gerbes 
simply gerbes. 0-gerbes are line bundles \cite{Hitchin}.} defined on a 
manifold $M$, in the pragmatic sense as in \cite{Hitchin} (see also 
\cite{Zunger}). This means that we define a gerbe by transition functions over 
patches, satisfying suitable cocycle identities. A connection on an $n$-gerbe 
is an  $(n+1)$--form\footnote{Although some may find the convention to label 
the gerbe by the number $n$ while it has a $(n+1)$--form connection confusing, 
it is actually convenient for discussions in physics. It implies that 
$n$-gerbes should naturally appear in the description of objects extended in 
$n$ spatial directions ($n$-branes), as these couple in a natural way to 
$(n+1)$--forms \cite{Zunger}.} $C_i$, defined over the patches $U_i$ of the 
manifold. On non-empty intersections $U_{ij} = U_i \cap U_j$ the forms over 
the patches $U_i$ and $U_j$ are related by a gauge transformation $C_i = C_j + 
\dif C_{ij}$. The $C_{ij}$ can be viewed as connections on $(n-1)$ gerbes 
defined at the overlap patches $U_{ij}$. To see why, note that as only 
$\dif C_{ij}$ is defined, the $C_{ij}$ are only defined up to closed 
forms (with integer periods), and hence have their own gauge invariance. On 
triple intersections $U_i \cap U_j \cap U_k $ there is the consistency 
condition $d(C_{ij} + C_{jk} + C_{ki}) = 0$ which is compatible with the 
ambiguity of adding closed forms. Therefore $C_{ij} + C_{jk} + C_{ki}$ and 
$C_{ij} + C_{jk} + C_{ki} + dC_{ijk}$ define the same cocycle on the triple 
overlap. The $C_{ijk}$ can be thought of as connections on an $(n-2)$-gerbe, 
defined over the $U_{ijk} = U_i \cap U_j \cap U_k$, which again have their 
own gauge invariance. This extends all the way until we arrive at zero-forms 
(functions), which can be thought of as transition functions for a 0-gerbe 
(that is, a line bundle).

When computing a specific holonomy, it is more convenient to reduce the
overlaps to infinitesimal size. Hence we cut up $M$ in pieces $M_i$ 
by a partition of unity. If the pieces $M_i$ and $M_j$ have a common 
boundary we denote this $M_{ij}$; a common boundary between $M_i$, $M_j$ 
and $M_k$ is written as $M_{ijk}$ etc. Define $(n+1)$-forms $C_i$ over 
the pieces $M_i$; one can think about the $M_{ij}$ as the places where one
``jumps'' from one patch to another (one can embed the $M_i$ in open 
patches $U_i$, then the $M_{ij}$ will be embedded in $U_{ij}$, etc. which 
gives our previous description). Again on overlaps, the connections in $M_i$ 
and $M_j$ are related by the gauge transformations $C_i = C_j + d C_{ij}$. 
$C_{ij}$ itself is defined over the $M_{ij}$, and has an intrinsic ambiguity 
by a closed form. One regards $C_{ij}$ as a connection on an $(n-1)$-gerbe 
over $M_{ij}$ with a $U(1)$ gauge invariance etc.

The concrete holonomy formula is (note the alternating sign):
\begin{eqnarray} \label{hol}
\int_M C = \sum_i \int_{M_i} C_i - \sum_{ij} \int_{M_{ij}} C_{ij} +
\sum_{ijk} \int_{M_{ijk}} C_{ijk} - \ldots.
\end{eqnarray}
This is invariant (for $M$ without boundary) under
\begin{eqnarray}
C_i  & \rightarrow & C_i + dL_i \nonumber \\
C_{ij} & \rightarrow & C_{ij} + L_i + L_j + d L_{ij} \nonumber \\
C_{ijk} & \rightarrow & C_{ijk} + L_{ij} + L_{jk} + L_{ki} + d L_{ijk}
\label{gt} \\
 & \ldots & \nonumber 
\end{eqnarray}
Two extreme cases of this formula are for a globally well-defined form
$C$, in which case the sum on the r.h.s. reduces to a single term, and
the case when only the last sum of integrals in the expression
contributes. These are analogues of the bundle case where physicists
are used to either using well-defined connections over large patches, or
to  ``putting the holonomy in the transition functions''. For the case of
connections on gerbes, there is a much larger freedom to ``put'' the
holonomy somewhere, due to the multiple gauge invariances in (\ref{gt}). 

Note that, even if the dimension of $M$ is smaller than the degree of the form 
$C$, eq. (\ref{hol}) gives a perfectly sensible, and not necessarily trivial 
result. One sets forms of too high degree formally to zero, but this does not 
necessarily imply that transition gerbes/bundles/functions are trivial. 
This may seem a rather pathological situation, but will be very useful 
to us below, because it actually allows us to avoid some technicalities 
when $M$ has a low dimension.

The above formula can be found in \cite{deBoer}, and was more or less 
inspired by gauge invariance, and inductive reasoning starting with the 
much simpler case of line bundles (0-gerbes). The reader may also want 
to consult \cite{Mackaay} and compare with eq.(8) and fig. 5 of that 
paper for the holonomy formula for 1-gerbes (or just gerbes for short).

In this paper we will use the above formula exclusively on $M= 
T^k/\Z_2$ (The fact that this $\Z_2$ defines an orientifold instead of an 
orbifold is relevant for the classification of charges, but not for the 
gerbe computations that follow). We choose coordinates $x_i$ $(i =1, 
\ldots,k)$ in $(\R/ 2\Z)^k$ on $T^k$, and quotient by the $\Z_2$ reflecting 
all coordinates. We label the $2^k$ fixed points of the $\Z_2$ by $x_i=p_i$, 
$p_i \in (\Z_2)^k$. The periodicity $2 \Z$ was chosen to allow us to work with 
the additive representation of the field of two elements, which we will denote 
as $\Z_2$. With this convention we can use ordinary addition and 
multiplication, with reduction modulo 2 understood, in a number of future 
computations. 

The $B$-field, which is a 2--form, can be regarded as a connection on a
(1-)gerbe.  We will require this gerbe to be flat, which means that 
the field strength $H= \dif B =0$. A non-zero $H$ would couple to 
gravity, and warp the background. We will compute the $B$-holonomy 
around an orientifold plane. Because the gerbe is flat, this is 
independent of the 2-surface we use to compute the holonomy (as long 
as it has the point $p$ in its interior, and the only possible source for 
$B$ is at $p$). For two different surfaces that can be deformed into 
eachother, the difference in holonomy is expressible as the integral of $H$ 
over the 3-surface that is swept out in the deformation, which has the two 
surfaces as boundary, and as $H$ is zero, this difference is 0. With the 
conventions used, it is possible to write down relatively simple expressions 
for gerbe holonomy defined over hypersurfaces around the $\Z_2$ fixed points.  

We pick a fixed point which we will denote by $p$, with coordinates 
$p_i$.  We define a box around this point by using the coordinate ranges  
$ p_{n-2} < x_{n-2} < p_{n-2} + \e,  \quad p_i - \e < x_i < p_i + \e $  
for $i=(n-1),n$, and localizing the box in the remaining directions by 
setting $x_i= p_i$ for $i < (n-2)$. The parameter $\e$ is not necessarily 
small; we only require that the box contains at most one possible source, and 
therefore $0 < \e < 1$. The surface of this box is a 2-surface; we will 
compute the $B$-field holonomy over it. This seems to assume that $n \geq 3$, 
but as noted before, this is not strictly necessary. For $n < 3$, we can still 
evaluate (\ref{hol}), but one truncates the computation by setting forms of 
too high degree, and the integrals with them formally to 0. We will not 
attempt to prove in all rigor that this procedure makes sense, but just note 
that this pragmatic approach reproduces known results.  

Let there be a 2-form field 
\bd
\sum_{i<j<k}b_{ij} \dif x_i \dif x_j/4
\ed 
in the bulk. We can assume that this is a constant 2--form; one can 
always use gauge transformations to set the 2-form locally to a 
constant. Another viewpoint is that this 2--form is inherited 
from $T^k$, as it is invariant under the $\Z_2$ action that we wish to 
quotient out, and hence survives the quotient. Because we can take it to 
be constant on $T^4$,  we can take it constant on the quotient. This fixes the 
first of the gauge invariances in (\ref{gt}), but it is not hard to see 
that the result of our computation will be gauge independent. We will 
integrate this over a surface that has the topology of $S^2/\Z_2= \R \p^2$. 

From now on we will always assume that the summed over indices are 
ordered. In the equations below, this will actually present a choice of 
gauge that is convenient for the choice of surface that we made. The 
first integral in the holonomy formula is over the faces of the cube; 
there are 5 of these, of which 4 pair up as opposite faces. The opposite 
faces do not contribute to the integral, because their contributions 
cancel (and the form is constant). The only contribution comes from the 
face at $x_{n-2} = p_{n-2} + \e $.  This gives 
\begin{equation}  
\label{b1} \sum_i \int_{M_i} B_i = b_{n-1,n} \e ^2 
\end{equation}

We next consider the transition functions: these are defined at $x_i =  p_i$ 
for $i \leq (n-2$), $  p_{n-1} < x_{n-1} <   p_{n-1} - \e $, $  p_n - \e < x_n 
<   p_n + \e $. The constant 2-form jumps upon traversing the plane at 
$x_{n-2} = p_{n-2}$ by an amount $b_{ij} \dif x_i \dif x_j/2$, and we should 
write this as the derivative of something. This is inherently ambiguous (by 
(\ref{gt})), so we fix the gauge and write 
\begin{equation}
b_{ij} \dif x_i \dif x_j = \dif \left( \sum_{i<j} b_{ij} x_i \dif x_j + 
b_{i} \dif x_i \right). 
\end{equation}
Here the second term is a closed 1-form that does not contribute to the
transition function, but will contribute to the holonomy. Independence 
of the gerbe from local definitions requires that the $b_i$ are independent of 
the patch we choose.

We have to integrate this 1-form over the edge of $x_i= p_i$ ($i \leq 
(n-2)$), $ p_{n-1} < x_{n-1} < p_{n-1} + \e $, $ p_n - \e < x_n < p_n + \e $.  
Again only one side contributes (the one with $x_{n-2} = p_{n-2}$, $x_{n-1} =
p_{n-1} + \e $), resulting in
\begin{equation} 
\label{b2}
\left( \sum_i b_{in} p_i + b_n \right) \e +b_{n-1,n}  \e^2
\end{equation}

The last contribution comes from the point at $x_i = p_i$ ($i \leq (n-1)$), 
$x_n = p_n + \e$. The transition function is \begin{eqnarray}
\left(\sum_{i<j} b_{ij} x_i x_j  + \sum_{i} b_{i} x_i + b \right)/2.
\end{eqnarray}
Again the value of $b$ cannot depend on the patch we 
are in. Inserting values for the coordinates gives
\begin{eqnarray} 
\sum_{i < j} b_{ij} p_i p_j +  \sum_i b_i p_i + 
b + \left(\sum_i b_{in} p_i  + b_n \right) \e. \label{b3} 
\end{eqnarray}

To find the total holonomy we add the contributions (\ref{b1}), 
(\ref{b2}), and (\ref{b3}). Because the gerbe is flat, we can actually 
replace our specific surface by any surface $M_p$ with the point $p$ (and no 
other source) in its interior. The final result is then: 
\be \label{bform} 
{\cal B}(\{p_i \})  \equiv \int_{M_p} B = \sum_{i < j} b_{ij} p_i p_j + 
\sum_i b_i p_i + b. 
\ee 
This formula is to be computed modulo 2; all coefficients $b_{ij}$, 
$b_i$, $b$ and $p_i$ are elements of $\Z_2$. The actual holonomy is 
\be
\exp i \pi {\cal B}( \{p_i \} )
\ee
and converts the additive representation of $\Z_2$ into its multiplicative one.
The holonomy is $1$ around an O$p^-$ or $\widetilde{\textrm{O}p}{}^-$, and 
$-1$ around a O$p^+$ or $\widetilde{\textrm{O}p}{}^+$. We see that, for 
sufficiently large $k$, the holonomy around each of the orientifold planes is 
completely specified by $k(k-1)/2$ coefficients of the antisymmetric tensor, 
$k$ coefficients $b_i$ and $1$ coefficient $b$. This is all the freedom one 
has, and as the NS charges are classified by the equivariant cohomology 
$H^2_{\Z_2}(T^k,\Z_2)$ \cite{deBoer} one should have 
\be 
H^2_{\Z_2}(T^k/\Z_2)= \Z_2^{\frac{k^2+k+2}{2}} \qquad k \geq 2 
\ee 
For $k=1$ we cannot define a 2--form, and one should have 
$H^2_{\Z_2}(S^1/\Z_2)= \Z_2^{2}$ because we can only choose $b_1$ and 
$b$. All this is in agreement with computations from 
\cite{deBoer}. The explicit generators in the appendix D of that paper 
correspond to a basis for the polynomials (\ref{bform}) (which form a 
vector space over the field $\Z_2$), by simply replacing the 
generator giving particular values at particular points by the 
polynomial associating the same values to those points. Finally, for 
$k=0$ one considers the equivariant cohomology of $S^0/\Z_2$. As $S^0$ is 
two points, $S^0/\Z_2$ is a single point, and one has the trivial result 
$H^2_{\Z_2}(S^0/\Z_2)= \Z_2$, which is correctly captured by formula 
(\ref{bform}) which in this case consists of the single coefficient $b$.

We note that for $k \geq 2$ we have $\frac{k^2+k+2}{2}$ parameters 
specifying the holonomies around $2^k$ points. For $k > 2$, the former 
quantity is smaller, and hence there will be relations among the planes. These 
can be described in terms of constraints. For $k=3$ one has 7 parameters 
describing 8 planes. The single constraint one has is easily derived from
\be 
\sum_{p_1=0}^1 \ \sum_{p_2=0}^1 \ \sum_{p_3=0}^1\left( \sum_{i < j} b_{ij} p_i 
p_j + \sum_i b_i p_i + b \right)=0. \ee
This tells us that summing the $\Z_2$ NS charges of all points gives zero, 
from which we conclude that the total number of NS charges is even. For 
$k=4$ one has 11 parameters for 16 planes, and one has 5 constraints. One of 
these tells again that the number of planes with NS charge is even. 
Furthermore, one may also fix a single coordinate and sum over the remaining 
ones, to see that the number of planes in every $T^3/\Z_2$ is even. There are 
4 independent choices for the fixed coordinate, and hence in total 5 
constraints. In lower dimensions, these constraints become increasingly 
difficult to analyze, and instead, we will use below a method which stays 
closer to the original formula's. 

Similar formula's can be derived for the RR $(5-p)$--form holonomy around an 
orientifold plane. In this case however, we have a different result 
for each dimension. We start with $T^k/\Z_2$ with $k=4$. The fixed 
points will be O$5$ planes, which can carry a RR 0--form charge. A 
zero-form is just a function, flatness requires this to be a constant 
function so we simply have
\be 
{\cal C}(\{p_i \}) = \int_{M_p} C  =  c  \label{cform4} 
\ee

For $k=5$ one has a 1--form RR charge. The relevant computation can be 
found in \cite{deBoer} (where it was computed for the orbifold K3 
$T^4/\Z_2$, but the computation here is completely analogous), and one 
finds 
\be 
{\cal C}(\{p_i \}) = \int_{M_p} C =  \sum_i c_i p_i + c \label{cform5}
\ee

For $k=6$ one has 2--forms, and we can copy our computation for the 
NS charges.
\be
{\cal C}(\{p_i \}) = \int_{M_p} C  =  \sum_{i < j} c_{ij} p_i p_j + \sum_i 
c_i p_i + c \label{cform6}
\ee 

For $k=7$ again the relevant formula can be found in \cite{deBoer}. 
\be
{\cal C}(\{p_i \}) = \int_{M_p} C =  \sum_{i < j 
< k} c_{ijk} p_i p_j p_k + \sum_{i < j} c_{ij} p_i p_j + \sum_i c_i p_i 
+ c  \label{cform7} 
\ee
The pattern will be clear by now.  The holonomy around the 
orientifold plane is $1$ around an O$p^-$ or O$p^+$, and $-1$ around a 
$\widetilde{\textrm{O}p}{}^-$ or $\widetilde{\textrm{O}p}{}^+$.

Therefore, the pair $({\cal B}( \{p_i \} ), {\cal C}( \{p_i \} ))$ completely 
determines the identity of the orientifold plane. We define
\bea
n_- & = & \{ \# p : {\cal B}( \{p_i \} )= 0, {\cal C}( \{p_i \} )=0 \} \\
n_+ & = & \{ \# p : {\cal B}( \{p_i \} )= 1, {\cal C}( \{p_i \} )=0 \} \\
\tilde{n}_- & = & \{ \# p : {\cal B}( \{p_i \} )= 0, {\cal C}( \{p_i \} )=1 \} \\
\tilde{n}_+ & = & \{ \# p : {\cal B}( \{p_i \} )= 1, {\cal C}( \{p_i \} )=1 \} 
\eea
The notation ``$\# p$'' should be read as ``the number of points $p$ with 
$\ldots''$. The numbers $n_-, n_+, \tilde{n}_-$ and $\tilde{n}_+$ count the 
numbers of  O$p^-$, O$p^+$, $\widetilde{\textrm{O}p}{}^-$ and 
$\widetilde{\textrm{O}p}{}^+$ respectively.

Another crucial number is represented by the RR tadpole. We represent this by 
the number $r$, defined as 
\be \label{tad}
r = 16 \ \frac{(n_- + \tilde{n}_-) - (n_+ + \tilde{n}_+)}{(n_- + 
\tilde{n}_-) + (n_+ + \tilde{n}_+)} - \frac{\tilde{n}_-}{2}
\ee
This formula expresses tadpole cancellation, where the number of 
D$p$-brane pairs is $r$. In sufficiently high dimension ($k < 
4$) $\tilde{n}_-$ and $\tilde{n}_+$ are actually zero, due to the 
absence of an RR charge for O$p$-planes in these dimensions. Of course 
$n_- + \tilde{n}_- + n_+ + \tilde{n}_+$ is simply the total number of 
fixed planes $2^k$, but we prefer to keep the sum explicit, as it 
clearly reveals the different ways in which the NS charge and RR charge 
for orientifold planes result in rank reduction.

For a consistent theory one has to cancel the RR tadpole. If $r$ is positive 
this can be done by adding $r$ pairs of D$p$ branes. If $r$ is negative, 
tadpole cancellation requires the addition of $r$ pairs of anti D$p$ branes, 
and this necessarily breaks supersymmetry. Below, we will restrict to $r$ 
non-negative, and we will refer to configurations with $r$ negative as 
``breaking supersymmetry''. If $r$ is non-negative, one may loosely refer to 
$r$ as the ``rank'' of the gauge group, although strictly speaking this is not 
true, as there are also gauge symmetries coming from Kaluza-Klein reduction on 
the metric and various forms. These however are not manifest in the 
orientifold set-up. 

By letting the coefficients $b$, $b_i$, $b_{ij}$, and $c$, $c_i$, $c_{ij}$, 
$c_{ijk}$ taking all possible values, one obtains all flat gerbes over 
$T^n/\Z_2$. This is too much; many of these gerbes can be related by 
using coordinate transformations on $T^n/\Z_2$. We will eliminate these 
redundancies by reducing the polynomials describing the gerbes to suitably 
chosen standard forms. Also, we will be less interested in the configurations 
that break supersymmetry (which in the case of a small number of non-compact 
dimensions outnumber the supersymmetric configurations), and therefore we will 
require $r \geq 0$.

The formula (\ref{bform}) is the same for all values of $k$. If one 
has a theory on $T^k/\Z_2$, with a particular configuration of 
NS charges, one can compactify the theory on (an) additional circle(s). 
If one does not turn on $B$-fields over these extra cycles, one can 
without difficulty T-dualize. The formula for the configuration on 
$T^{k+n}/\Z_2$ remains the same, which implies that the numbers $(n_- + 
\tilde{n}_-)$ and $(n_+ + \tilde{n}_+)$ are multiplied by powers 
of 2.

Upon using a similar procedure on the formula's for the RR charge, 
(\ref{cform4}) should be mapped to (\ref{cform5}) which in turn gets 
mapped to (\ref{cform6}), and so on. We know how the top-form (the 
coefficient in ${\cal C}$ with the most indices) transforms; it simply 
gains the index of the direction in which we dualize. It follows that 
the same must be true for all the coefficients in the polynomial ${\cal 
C}(\{p_i \})$; all gain an index upon T-dualizing, and the map of the 
polynomials into each other is now obvious. We also see that the number 
of planes with RR charge stays the same under T-duality. As a 
consequence of these rules, the formula (\ref{tad}) is invariant 
under T-duality (as it should be).

All this is perfectly consistent with the T-duality rules ($q =p+1$):
\bea
\textrm{O}q^- & \leftrightarrow & \textrm{O}p^- + \textrm{O}p^- \\
\textrm{O}q^+ & \leftrightarrow & \textrm{O}p^+ + \textrm{O}p^+ \\
\widetilde{\textrm{O}q}{}^- & \leftrightarrow & \textrm{O}p^- + 
\widetilde{\textrm{O}p}{}^- \\ 
\widetilde{\textrm{O}q}{}^+ & \leftrightarrow & \textrm{O}p^+ + 
\widetilde{\textrm{O}p}{}^+ \eea  

We conclude that the formalism we have set up is both self-consistent 
and consistent with known facts about orientifolds. We now turn to study 
the holonomy formula's, and derive their implications for the possible 
configurations of orientifold planes. 

 \section{Models with at most two kinds of planes} 
\label{twoplanes}

Before dealing with the general case, we first study the equations for 
${\cal B}(\{p_i \})$ and ${\cal C}(\{p_i\})$ separately. This amounts to 
studying configurations where only one of the $\Z_2$ charges, and hence 
at most two kinds of orientifold planes are present.

\subsection{Models with $\mathrm{O}p^-$ and $\mathrm{O}p^+$ only} \label{B}

We will start by first classifying all configurations of NS charges that 
give rise to flat gerbes. As these are reflected in ${\cal B}( \{p_i \} )$, we will 
set ${\cal C}( \{p_i \} )=0$ for the remainder of this subsection. Doing so 
immediately will give us a list of configurations with O$p^-$ and 
O$p^+$-planes that give rise to flat gerbes. We will still impose the 
requirement of supersymmetry. Combining $r \geq 0$ with formula 
(\ref{tad}), one sees that in this context this translates into $n_- 
\geq n_+$. At the end of the subsection we will add some comments on 
non-supersymmetric configurations.

We are only interested in configurations modulo the action of coordinate 
transformations on $T^k/\Z_2$. We therefore fix the polynomial 
${\cal B}( \{p_i \} )$ to a particular standard form by using these 
transformations.

On $T^k$, the symmetry group would be 
\bd
(\R/ 2\Z)^k \ltimes SL(k,\Z),
\ed 
with $(\R/2\Z)^k$ the group of translations $\R^k$ up to periodicity 
$2\Z^k$, and $Sl(2,\Z)$ the mapping class group of the $k$--torus. The 
symmetry group of the $\Z_2$ quotient is: 
\bd
\Z_2^k \ltimes SL(k,\Z_2)
\ed
The $\Z_2$ quotient breaks the translation group $(\R/2\Z)^k$ to $\Z_2^k$ 
because the periodicity is affected. The group $SL(k, \Z)$ is not broken, but 
elements of these group that have equal entries modulo $2$ have the same 
action on the $\Z_2$ quotient (as far as the orientifold planes are 
concerned), and therefore the group is projected to $SL(k, \Z_2)$.

The elements of the $\Z_2^k$ subgroup are generated by the affine 
transformations $x_i \rightarrow x_i +1$. The elements of $SL(k,\Z_2)$ are 
linear transformations (over the field $\Z_2$ one has 
$GL(k,\Z_2)=SL(k,\Z_2)$). Acting with these on the polynomial ${\cal B}( \{p_i 
\} )$, we will now bring it to a standardform.

Pick two coordinates $x_i$ and $x_j$ such that $b_{ij} = 1$. If there are no
such coordinates then the quadratic part of the polynomial is in standardform
already. If there are, then rename $x_i \leftrightarrow x_1$ and $x_j
\leftrightarrow x_2$. In the expression for ${\cal B}( \{p_i \} )$ there now 
are a number of tems that involve $p_1$. Write these as
\bd
p_1 \left(p_2 + \sum_{i \neq 2} b_{1i} p_i + b_1 \right)
\ed
Setting $x_2 \rightarrow \left(x_2 + \sum_{i \neq 2} b_{1i} x_i + b_1 \right)$ 
eliminates all of these terms except $p_1 p_2$. After this operation the only 
term in ${\cal B}( \{p_i \} )$ that depends on $p_1$ is $p_1 p_2$. One then 
isolates all terms containing $p_2$ which can be written as \bd p_2 
\left(p_1 + \sum_{i \neq 1} b_{2i} p_i +b_2\right) \ed
and sets $x_1 \rightarrow \left(x_1 + \sum_{i \neq 1} b_{2i} x_i +b_2 
\right)$. The only term left that still contains $p_1$ or $p_2$ is the 
product $p_1 p_2$. One then repeats the procedure with $p_3$ and $p_4$ etc. 
until done.

We now have transformed to a configuration with only $b_{i, i+1}$ ($i$ odd)  
possibly non-zero, and if $b_{i, i+1} = 0$, then so are all $b_{j, j+1} = 0$ 
for $j \geq i$. If $m$ is the smallest odd integer for which $b_{m,m+1}=0$
then also all $b_i=0$ for $i < m$. If any of the $b_i$ with $i \geq m$ is 
non-zero we first interchange (if necessary) $x_i$ with $x_m$, to set 
$b_m=1$. Subsequently one transforms
\bd
x_m \rightarrow \sum_i b_i x_i + b
\ed
to set all $b_i$ with $i > m$, and $b$ to zero.

With these manipulations we have arrived at what we will use as 
standardform. The standardform is specified by a number of parameters $b_{i, 
i+1}$, $i$ odd, $i < m$ for some odd $m$, a parameter $b_m$, and a 
parameter $b$ (which can only be non-zero if $b_m=0$ ). These completely 
specify the NS charges of a given orientifold configuration on $T^k/\Z_2$, up 
to coordinate transformation (linear or affine). 

It is useful to relax for a moment the constraint that $b$ can only be 
non-zero if $b_m=0$. Doing this, we note an elegant symmetry: replacing $b$ by 
$b+1$, one exchanges O$p$-planes with NS charge, with planes without 
NS charge. Therefore this symmetry exchanges $n_-$ and $n_+$. When $b_m=1$, 
the transformation $x_m \rightarrow x_m+1$ precisely has the effect $b 
\rightarrow b+1$. In this case $b \rightarrow b+1$ is an automorphism of the 
configuration of orientifold planes, and this implies that if $b_m = 1$, $n_+ 
=n_-$. This in turn implies that $r=0$ for these models.

On the other hand, if $b_m =0$, the action of $b \rightarrow b+1$ does not 
correspond to any coordinate symmetry. In this case $b \rightarrow b+1$ has 
the effect $r \rightarrow -r$. Hence, only one of the configurations can be 
supersymmetric, and it is not hard to verify that, with the chosen 
standardform, this is always the one with $b=0$.

Consequently, the requirement of supersymmetry together with the chosen 
standardform, leads us to always set $b=0$.

Putting all information together, one arrives at table 
\ref{NScharges}. The first column of this table lists the maximal dimension of 
non compact space $d_{max}$. If a certain model exists in $d_{max}$, one 
can further compactify it on an additional $n$--torus, and use T-dualities to 
deduce the existence of lower dimensional models. The subsequent column lists 
the non-zero coefficients of the polynomial ${\cal B}( \{p_i \} )$. This 
specifies the model completely, and determines the number of various kinds of 
O$p$ planes, and their location. For easy reference, the numbers $n_-$ and 
$n_+$ that follow from ${\cal B}( \{p_i \} )$ are given in the next columns. 
These are functions of the parameter $k$ appearing in $T^k/\Z_2$. Of course 
one should remember that always $k \geq 9-d_{max}$. The last column lists the 
quantity $r$, which is independent of $k$, as it should be.

\TABLE{ 
\begin{tabular}{|c||cc|cc||c|} \hline 
$d_{max}$  &      &    & $n_-$ & $n_+$ & $r$ \\  \hline
\hline 
9  & $b_{ij} = 0$ & $b_i=0$  &   $2^k$         &    0        & 16 \\
   
8  & $b_{ij} = 0$ & $b_1=1$  &   $2^{k-1}$        &    $2^{k-1}$        & 0 \\

7  & $b_{12} =1$ & $b_i=0$  &   $3 \cdot 2^{k-2}$       &  $2^{k-2}$  & 8 \\

6  & $b_{12} = 1$ & $b_3=1$ &   $2^{k-1}$      &    $2^{k-1}$       & 0  \\

5  & $b_{12} = b_{34}=1$ & $b_i=0$ & $5 \cdot 2^{k-3}$ & $3 \cdot 2^{k-3}$ & 4 
\\

4  & $b_{12} = b_{34}=1$ & $b_5=1$   & $2^{k-1}$ & $2^{k-1}$ & 0 \\

3  & $b_{12} = b_{34}= b_{56}=1$ & $b_i=0$   & $9 \cdot 2^{k-4}$ & $7 \cdot 
2^{k-4}$ & 2 \\

2  & $b_{12} = b_{34}= b_{56}=1$ & $b_7=1$   & $2^{k-1}$ & $2^{k-1}$ & 0 \\

1   & $b_{12} = b_{34}= b_{56}= b_{78}=1$ & $b_i=0$  & $17 \cdot 2^{k-5}$ & 
$15 \cdot 2^{k-5}$ & 1 \\

0   & $b_{12} = b_{34}= b_{56}= b_{78}=1$ & $b_9=1$  & $2^{k-1}$ & $2^{k-1}$ & 
0 \\

\hline
\end{tabular} 
\caption{$T^k/\Z_2$ orientifolds with O$p^+$ and O$p^-$ planes only.}  
\label{NScharges} } 

The entries in table \ref{NScharges} exhibits a very simple structure, that is 
actually easy to understand. First of all, in every dimension exactly one 
new model appears, and therefore there are $k+1$ possible models for any value 
of $k$. The models in table \ref{NScharges} can be divided in two classes, 
depending on the value of $b_m$. For $b_m=1$, the models have their ranks 
reduced by powers of 2. The first model in this chain, $b_{ij}=0$ with rank 
16, is just the T-dual of the standard toroidal compactification of type I 
theory, with its $Spin(32)/\Z_2$ gauge group . The other 
elements in the chain are T-dual to the various possibilities of compactifying 
type I theory without vector structure \cite{Bianchi, Bianchi2, Witten97, 
Angelantonj}. The values for $b_{ij}$ in the table are also the values for the 
$B$--field on the $k$--torus in the T-dual type I theory.

The class of models with $b_m=1$ has equal numbers of O$p^+$ and O$p^-$ 
planes. The simplest of these is the model with $d_{max}=8$. It has one 
O$8^+$ and one O$8^-$ plane, and is dual to type IIB on $S^1/\delta \Omega$. 
This is an orientifold, but here the worldsheet parity $\Omega$ is combined 
with a translation $\delta$ over half the circumference of the circle $S^1$ 
\cite{Dabholkar, Witten97}. Also the other models with $b_m=1$ (and therefore 
$r=0$) can be understood this way. The existence of these models, which have 
the same number of various orientifold planes, but inequivalent geometries, 
was first noted in \cite{Bergman, deBoer}. These appear at even values for 
$d_{max}$, and can be understood as duals to compactifications of type IIB on 
$S^1/\delta \Omega \times T^n$, where there are additional $B$-fields turned 
on on $T^n$ \cite{Keurentjes01}.

Incidentally, we note that, as promised, in low compact dimension 
($9 \geq p \geq 6$) our formalism reproduces the known orientifolds 
\cite{Bergman, deBoer, Witten97}, even though the original 
definition of ${\cal B}$ as the integral over an $\R\p^2$ surrounding the O$p$ 
plane runs into difficulties here. 

Before concluding this section, we briefly return to the class of
non-supersymmetric models we discarded, the ones with $b_m=0, 
b=1$. We start with the simplest of these models, described by $b=1$ and 
all other coefficients 0. This would be a 10-dimensional theory with 
$Sp(16)$ group, realized by composing 32 anti-branes with an 
O$9^+$-plane \cite{Sugimoto}, and its T-duals. Also $Sp(16)$ bundles on a 
$k$-torus allow bundles with twisted boundary conditions, and had we carried 
out our programme while neglecting supersymmetry, we would have found the 
T-dual descriptions for all of these. The list would simply be the same 
list as for the type I duals, but with an extra $b=1$, and the numbers 
$n_-$ and $n_+$ interchanged. A last caveat: these models can only be 
interpreted as representing the flat connections in a classical gauge 
theory. In absence of supersymmetry, it should be expected that the 
degeneracy of vacua will be lifted by quantum corrections. The anti 
D$p$-branes break supersymmetry and in particular the total configuration is 
not a BPS-system. As a consequence, it should be expected that a 
generic configuration is unstable, and the anti D$p$ branes are dynamically 
driven to special locations. This looks like a T-dual description of the 
localization of the wave function that is well known for finite volume 
non-supersymmetric gauge theories (see e.g. \cite{vanBaal} and references 
therein). Actually, even this final state may be only an approximate one, as 
it is not unlikely that such a theory may tunnel by non-perturbative effects 
into a stabler (supersymmetric?) minimum. 

\subsection{Models with $\mathrm{O}p^-$ and $\widetilde{\mathrm{O}p}{}^-$ 
only}  \label{C} 

We now turn to classification of flat gerbe coming from RR charges only, 
meaning that in the remainder of this section we will set ${\cal B}( \{p_i \} 
)$ to zero. This automatically gives us the classification of orientifolds 
$T^k/\Z_2$ with O$p^-$ and $\widetilde{\textrm{O}p}{}^-$ only. The polynomial 
${\cal C}( \{p_i \} )$ is different for any value of $k$, and we cannot treat 
all dimensions simultaneously, like in the previous section.

For $k < 4$ there is no formula for ${\cal C}( \{p_i \} )$ because in these 
models the O$p$-planes cannot be given an RR charge.

In $k=4$ one has the simple equation (\ref{cform4}) which says ${\cal C}( 
\{p_i \} ) = c$. This equation is invariant under all reparametrizations of 
$T^k/\Z_2$, and therefore automatically in ``standardform''. Setting $c=0$ 
reproduces the standard compactification, with 16 O$5^-$ planes. Setting $c=1$ 
turns all into $\widetilde{\textrm{O}5}{}^-$.

In $k=5$ one has formula (\ref{cform5}), which is the simple linear equation 
${\cal C}( \{p_i \} ) = \sum_i c_i p_i + c$. We first have to bring this to 
standardform. If at least one of the $c_i$ is non-zero, one uses, if necessary 
$x_1 \leftrightarrow x_i$ to set $c_1=1$. Subsequently one transforms $x_1 
\rightarrow {\cal C}(\{x_i\})$ (by which we mean ${\cal C}(\{p_i\})$, 
with $p_i$ replaced by $x_i$), and ends up with $c_1=1$ and all other 
coefficients zero.  Upon using a T-duality in the $1$-direction, we see 
that this is actually T-dual to a configuration that appeared for $k=4$. 
The only remaining option is to set all $c_i=0$, and set $c=1$. As in 
$k=4$, this is invariant under any coordinate transformation. It results 
in a configuration with 32 $\widetilde{\textrm{O}4}{}^-$ planes.

The one-form in $k=5$ can be interpreted as a connection on a bundle. In 
this case we have IIA theory and of course in this case the total space 
of the bundle is interpreted as the manifold upon which M-theory is 
compactified. Orientifold planes with RR 1--form charge correspond to 
twists of this bundle \cite{Gimon98, Hori}, and hence the classification 
of bundles is the same as the classification of those configurations 
that can be lifted to M-theory (see the remarks in 
section 3.2.6 of \cite{deBoer}).

In $k=6$ we have equation (\ref{cform6}). This is the same formula as for the 
$B$-fields, and one can use the tricks of the previous section to rewrite any 
configuration into a standard form. Again $c=0$, because of the tadpole 
requirement. The configurations $c_1 =1$, and the configuration $c_{12} =1$ 
are related by T-dualities to previously considered models. The other ones are 
new ones.

For $k > 6$ our previous methods fail. Essentially this is because we do not 
know how to rewrite the tri-linear term in (\ref{cform7}) or higher order 
terms appearing for still higher values of $k$ to a standard form. Because 
of this technical problem our classification of configurations with RR charges 
will not proceed beyond $k=6$.

\TABLE{ 
\begin{tabular}{|c||cc|cc||c|} \hline 
$d_{max}$  &  &  & $n_-$ & $\tilde{n}_-$ & $r$ \\ 
\hline
\hline 
5  & $c_{[5..k]}=0$ &   &    $2^k$        & 0  &  16\\
    & $c_{[5..k]}=1$ &   &    $2^k -16$          & 16 & 8 \\
\hline
4  & $c_i=0$ & $c_{[6..k]}=1$  & $2^k-32$ & 32 & 0 \\
\hline
3  & $c_{12[7..k]} = 1$ & $c_{3[7..k]}=1$   &   $2^k-32$      &   32      & 0  
\\     & $c_{12[7..k]} = c_{34[7..k]}=1$ & $c_{i[7..k]}=0$ &  $2^k-24$ & 24 & 
4 \\     & $c_{12[7..k]} = c_{34[7..k]}=1$ & $c_{5[7..k]}=1$   & $2^k-32$ & 32 
& 0 \\     & $c_{12[7..k]} = c_{34[7..k]}= c_{56[7..k]}=1$ & $c_{i[7..k]}=0$ & 
$2^k-28$ & 28 & 2 \\   

\hline
\end{tabular} 
\caption{$T^k/\Z_2$ orientifolds with $\widetilde{\textrm{O}p}{}^-$ and 
O$p^-$ planes only.}\label{RRcharges} }  

Our results of this section are compiled in table \ref{RRcharges}. The table 
has the same structure as table \ref{NScharges} except that now O$p^+$ 
planes, and therefore $n_+$ are absent. Instead 
$\widetilde{\textrm{O}p}{}^-$ planes appear, with their associated quantity 
$\tilde{n}_-$. The only other new element in table \ref{RRcharges} is the 
notation we introduced in the coefficients of ${\cal C}( \{p_i \} )$. As 
explained previously, these coefficients gain indices when T-dualizing. We 
tried to capture this in the notation $[m..k]$. The idea is that this 
represents a string of subsequent indices ranging from $m$ to $k$. If $k < m$ 
it represents an empty string. As an example, the notation $c_{12[7..k]}$ 
stands for $c_{12}$ if $k=6$, for $c_{127}$ if $k=7$, for $c_{1278}$ if $k=8$, 
etcetera.

The entries in this table have a much more irregular structure than the 
previous table. The first non-trivial theory in table \ref{RRcharges} (the 
second entry) is T-dual to a type I compactification on $T^4$ with a so-called 
non-trivial quadruple of holonomies \cite{deBoer, Kac, Keurentjes00b}. These 
belong to a class of only fairly recently discovered compactifications of 
non-Abelian gauge theories on tori, having flat bundles that are not connected 
via flat connections to the trivial flat connection, but are also not 
characterized by their topology.

The entry under $d_{max} = 4$ is another example of such a compactification, 
this time T-dual to a type I compactification on $T^5$ with a non-trivial 
quintuple of holonomies \cite{deBoer, Kac}. The interpretation of the theories 
listed under $d_{max}=3$ is less straightforward, although their holonomies 
are fairly easy to write down. As explained in section 5 of 
\cite{Keurentjes00b}, they can be obtained by ``superposing'' various 
quadruple configurations, and reducing D$p$ branes modulo 2.

Besides flat gauge bundles with quadruples and quintuples, there also exist 
flat gauge bundles with triples. These last bundles were thoroughly studied in a number of 
papers \cite{Borel, Kac, Keurentjes99}. Unfortunately, there are severe 
constraints on the existence of such compactifications in string theory, and 
as a matter of fact non-trivial triples do not give rise to flat gauge bundles 
in $Spin(32)/\Z_2$ string theory \cite{deBoer}. Although non-trivial flat 
bundles on higher dimensional tori can be realized, the systematics for these 
theories is not completely understood yet from the gauge theory side. 

We conclude this section with a few remarks on the  non-supersymmetric 
models that were discarded. With the interpretation of an 
$\widetilde{\textrm{O}p}{}^-$ plane as a bound state of an O$p^-$ plane with a 
D$p$-brane, and the simple formula for $r$ for the theories in this 
section 
\be 
r = 16 - \frac{\tilde{n}_-}{2}
\ee
one may interpret these as theories that fail to be supersymmetric because of 
an insufficient number of D$p$ branes in the theory. Indeed, were it not for 
tadpole cancellation, these theories would make perfect sense as dual 
descriptions to compactifications of $O(n > 32)$ open string theories. With 
the requirement of tadpole cancellation, one is automatically forced to 
introduce $-r$ pairs of anti D$p$ branes. A pair of anti D$p$ branes will feel 
an attractive force towards an $\widetilde{\textrm{O}p^-}$ plane, and having 
arrived at this plane should annihilate with the D$p$ brane there to form a 
bound state of an O$p^-$ plane with an anti D$p$ brane. Hence these theories 
will finally have $(\tilde{n}_- + r)$ $\widetilde{\textrm{O}p}{}^-$ planes 
(remember that $r$ is negative for non-supersymmetric theories) and $-r$ 
(non-supersymmetric) bound states of O$p^-$ with a single anti D$p$ on 
top. 

The positions of these planes are not fixed; by D$p$ brane-antibrane pair 
creation in the bulk and letting such pairs annihilate with D$p$ or anti 
D$p$'s at the orientifold planes one can permute $\widetilde{\textrm{O}p}{}^-$ 
planes with a bound state of O$p^-$ with anti D$p$. Note however that these 
processes can never convert an O$p^-$ into an $\widetilde{\textrm{O}p}{}^-$ or 
the bound state of O$p^-$ and anti D$p$. Again, in the absence of 
supersymmetry, the degeneracy in vacuum configurations is not protected; the 
actual vacuum should be a superposition of all possible semiclassical vacua. 
Again we are assuming here that this vacuum is ``sufficiently'' stable, it is 
conceivable that by non-perturbative effects this theory may tunnel to another 
vacuum which is not at all resembling this orientifold description. However, 
under suitable circumstances the time-scale for such a process may be much 
longer than the time scale that the orientifold would need to find its 
metastable vacuum.

\section{Mixed cases: All kinds of planes} \label{BC}

Finally we will tackle the general case, where both ${\cal B}( \{p_i \} )$ and 
${\cal C}( \{p_i \} )$ can be non-zero polynomials. We will 
omit the cases ${\cal B}( \{ p_i \})=0$ and ${\cal C}( \{p_i \} )=0$ as their 
descriptions can be found in the previous sections. Because of the problems 
with ${\cal C}( \{p_i \} )$ for $k > 6$, our classification will terminate at 
$k=6$. In $k=6$ the computations are very lengthy, although relatively 
straightforward. For the ease of the reader we will first present and discuss 
our results, and present the justifying computations later. 

\TABLE{
\begin{tabular}{|c||c|cccc||c|} \hline 
$d_{max}$  &    & $n_-$ & $n_+$ & $\tilde{n}_-$ & $\tilde{n}_+$ & $r$ \\ 
\hline
\hline 
5  & $b_{12} =1$  &  & & & & \\
    & $c_{[5..k]}=1$ & $3 \cdot 2^{k-2}-12$ & $2^{k-2}-4$ & 12 & 4 & 2 \\
\hline \hline 
4 &  $b_1=1$ & & & & &\\
& $c_{1[6..k]}=1$ & $2^{k-1}$ & $2^{k-1}-16$ & 0 & 16 & $0_a$ \\  \hline
&  $b_{12} = 1$ & & & & &\\
& $c_{1[6..k]} =1$ & $3 \cdot 2^{k-2} -8 $ & $2^{k-2}-8$ & 8 & 8 & 4 \\ \hline
    &  $b_{12} = 1$ & & & & & \\
& $c_{1[6..k]} = c_{[6..k]}=1$ & $3 \cdot 2^{k-2} -16 $ &  $2^{k-2}$ & 16 & 0 
& $0_b$ \\   \hline
  &  $b_{12}=b_{34} =1$ & & & & & \\
&  $c_{1[6..k]} =1$ & $5 \cdot 2^{k-3} -8$ & $3 \cdot 2^{k-3} -8$ & 8 & 8 & 
$0_c$ \\   \hline 
\end{tabular} 
\caption{$T^k/\Z_2$ orientifolds with all kinds of planes, $d_{max} > 3$}  
\label{mixeda} }

\TABLE{ 
\begin{tabular}{|c||c|cccc||c|} \hline 
$d_{max}$  &    & $n_-$ & $n_+$ & $\tilde{n}_-$ & $\tilde{n}_+$ & $r$ \\ 
\hline \hline 3 & $b_1 =1$ & & & & & \\
& $c_{1[7..k]} =1$  & $2^{k-1}$ & $2^{k-1}-32$ & 0 & 32 & 0 \\ \hline
 & $b_{12} =1$ & & & & & \\
& $c_{12[7..k]} =c_{34[7..k]}=1$ & $3 \cdot 2^{k-2}-12$ & $2^{k-2}-12$ & 12 & 
12 & 2 \\ \hline
 & $b_{12} =1$ & & & & & \\
 & $c_{1[7..k]} =1$ & $3 \cdot 2^{k-2} -16$ & $2^{k-2} -16$ & 16 & 16 & $0_a$ 
\\ \hline
 & $b_{12} =1$ & & & & & \\
 & $c_{12[7..k]} =1$ & $3 \cdot 2^{k-2}$ & $2^{k-2} -16$ & 0 & 16 & 8 \\
\hline
 & $b_{12} =1$ & & & & & \\
 & $c_{12[7..k]} = c_{1[7..k]}=1$ & $3 \cdot 2^{k-2}-16$ & $2^{k-2}$ & 16 & 0 
& $0_a$ \\ \hline 
 & $b_{12} =1$ & & & & & \\
 & $ c_{15[7..k]}= c_{34[7..k]}=1$ &  $3 \cdot 2^{k-2}-16$ & $2^{k-2}-8$ & 16 
& 8 & $0_c$ \\ \hline 
 & $b_{12} =1$ & & & & & \\
 & $c_{13[7..k]}= c_{2[7..k]}=c_{3[7..k]}= 1$ &  $3 \cdot 2^{k-2}-16$ & 
$2^{k-2}-16$ & 16 & 16 & $0_b$ \\ \hline 
 & $b_{12} =1$ & & & & & \\
 & $ c_{14[7..k]}= c_{23[7..k]}=1$ &  $3 \cdot 2^{k-2}-16$ & $2^{k-2}-8$ & 16 
& 8 & $0_d$ \\ \hline
 & $b_{12} = b_3 =1$ & & & & & \\
 & $c_{12[7..k]} = c_{23[7..k]} = 1$ & $2^{k-1}$ & $2^{k-1} - 16$ & 0 & 16 & 
$0_b$ \\ \hline
 & $b_{12} = b_3 =1$ & & & & & \\
& $c_{12[7..k]} = c_{3[7..k]} = 1$ & $2^{k-1}$ & $2^{k-1} - 32$ & 0 & 32 & 0 
\\ \hline
 & $b_{12} = b_{34}=1$ & & & & & \\
 & $c_{12[7..k]} = 1$ & $5 \cdot 2^{k-3}-4$ & $3 \cdot 2^{k-3}-12$ & 4 & 12 & 
2 \\ \hline 
 & $b_{12} = b_{34}=1$ & & & & & \\
 & $c_{13[7..k]} = 1$ & $5 \cdot 2^{k-3}-8$ & $3 \cdot 2^{k-3}-8$ & 8 & 8 & 
$0_d$ \\ \hline  
& $b_{12} = b_{34}=1$ & & & & & \\
 & $c_{13[7..k]}=c_{24[7..k]}=1$  & $5 \cdot 2^{k-3}-8$ & $3 \cdot 2^{k-3}-16$ 
& 8 & 16 & $0_d$ \\ \hline
 & $b_{12} = b_{34}=1$ & & & & & \\
 & $c_{12[7..k]} = c_{34[7..k]}=1$ & $5 \cdot 2^{k-3}$ & $3 \cdot 2^{k-3}-24$ 
& 0 & 24 & 4 \\ \hline
 & $b_{12} = b_{34}=1$ & & & & & \\
 & $c_{15[7..k]}=c_{34[7..k]}=1$ & $5 \cdot 2^{k-3}-8$ & $3 \cdot 2^{k-3}-8$ & 
8 & 8 & $0_c$ \\ \hline 
 & $b_{12} = b_{34} =b_5 =1$ & & & & & \\
& $c_{12[7..k]} = c_{34[7..k]} = c_{5[7..k]} = 1$ & $2^{k-1}$ & $2^{k-1} - 32$
& 0 & 32 & 0 \\  \hline
 & $b_{12} = b_{34} =b_{56} =1$ & & & & & \\
& $c_{12[7..k]} = c_{34[7..k]} = c_{56[7..k]} = 1$ & $9 \cdot 2^{k-4}$ & 
$7 \cdot 2^{k-4} - 28$ & 0 & 28 & 2 \\  \hline
\end{tabular} 
\caption{$T^k/\Z_2$ orientifolds with all kinds of planes, 
$d_{max}=3$} \label{mixedb}  } 

Our results can be found in the tables \ref{mixeda} and \ref{mixedb}. These 
tables have the same structure as the tables \ref{NScharges} and 
\ref{RRcharges}, except that now one has 4 columns to count all the kinds of 
orientifold planes. We will explain in section \ref{S} why we have attached an 
index to some of the zero's in the last column. 

The theories described in the tables \ref{mixeda} and \ref{mixedb} are 
``mixed'' in multiple ways. First of all, ``mixed'' refers to the presence of 
both NS and RR charges in the characterization of the orientifold planes. But 
they are also mixed in the sense that their dual description involves a 
mixture of the elements of the dual descriptions in the previous sections.

First consider the models where less than half of the O$p$ planes carries an 
NS charge. As explained in section \ref{B} such models are T-dual to type I 
theories without vector structure. In section \ref{C} we explained that for
topologically trivial, flat $Spin(32)/\Z_2$ bundles there are components in 
the gauge theory moduli space that are not connected to the trivial flat 
bundle via flat connections, and that these give rise to new orientifold 
models. It should come as no surprise that also for topologically non-trivial 
bundles there are many different components in the moduli space of flat 
bundles \cite{Borel}, and that these give rise to different orientifold 
descriptions \cite{deBoer, Keurentjes00a}. 

The models where half of the O$p$-planes carry NS charges can still be 
identified with duals to the type IIB-theory on $S^1/\delta \Omega$. Because 
in this case $n_- + \tilde{n}_- = n_+ + \tilde{n}_+$, a glance at formula 
(\ref{tad}) will teach us supersymmetry requires that $\tilde{n}_- =0$. This 
is also intuitively clear; as the tadpole is already cancelled because of the 
balance between planes with and without NS charge, there can be no excess 
D$p$-branes forming bound states with O$p^-$ planes.

This argument does not rule out the presence of 
$\widetilde{\textrm{O}p}{}^+$ planes in these theories, and indeed various 
solutions with these planes exist. It is however not easy to describe 
accurately the distinction between these theories, and the ones where one 
replaces the $\widetilde{\textrm{O}p}{}^+$ planes by O$p^+$-planes. Both are 
duals of the type IIB-theory on $S^1/\delta \Omega$, and their difference is 
due to rather subtle RR phases on the torus upon which the theory is 
compactified, which are rather poorly understood at present.

Also for the (large number of) non-supersymmetric theories that are discarded 
there, it is hard to make general statements. In principle the starting point 
is again clear: when one computes $r$ to be negative, one adds anti D$p$ 
branes to cancel the tadpole. Anti D$p$ branes can annihilate with D$p$ 
branes at $\widetilde{\textrm{O}p}{}^-$ planes, and the endpoint of the 
whole process is depending on $r$ and $\tilde{n}_-$. If $-r > \tilde{n}_-$ 
there will be free anti D$p$ branes left. These will then probably be driven 
to some (meta)stable non-supersymmetric configuration of minimal vacuum 
energy. If $-r < \tilde{n}_-$ all D$p$ branes at 
$\widetilde{\textrm{O}p}{}^-$ planes will be annihilated, and one has a 
configuration where all D$p$  and anti D$p$ branes form bound states with 
O$p^-$ planes, and the true vacuum will be a superposition of all possible 
configurations. Finally if $-r = \tilde{n}_-$, all anti D$p$ branes will 
annihilate, and the final state is, at least semiclassically, 
unique\footnote{An example of such an orientifold is $T^6/\Z_2$, with ${\cal 
B}=p_1 p_2 +1$, and ${\cal C}=p_1 p_2$. This demonstrates that this set of 
models is not empty.}.

We will now turn to the actual computations that lead to the tables 
\ref{mixeda} and \ref{mixedb}. We start with the easiest computation, for $k = 
4$. If at any place the reader gets weary of the computations, he is 
encouraged to skip the remainder of this section and turn to section \ref{S}.

\subsection{5+1 dimensional theories}

Our general line of attack in this subsection and the following ones will 
be as follows. First we fix a configuration of NS charges. This amounts 
to picking an entry from table \ref{NScharges}, and inserting the 
coefficients in the polynomial ${\cal B}( \{p_i \} )$. After having fixed this 
we consider in all generality all possibilities for ${\cal C}( \{p_i \} )$. We 
have to do so, because the fact that we have already used part of the 
symmetries to fix ${\cal B}( \{p_i \} )$ means that we have less freedom to 
manipulate ${\cal C}( \{p_i \} )$. 

It will be clear that there are two factors determining the difficulty of 
this process: 1) When we will go to lower dimensions (higher $k$) the 
polynomial ${\cal C}( \{p_i \} )$ will become a more complicated expression; 
and 2) the more coefficients of ${\cal B}( \{p_i \} )$ are fixed, the fewer 
symmetries we have at our disposal. This is why the first computations we will 
present are almost trivial, while by the end of the third subsection they will 
have grown very tedious. In these subsections, when we have found a 
supersymmetric model we will compute the row of 5 numbers $(n_-, n_+, 
\tilde{n}_-, \tilde{n}_+, r)$ and list it. 

Let us apply the procedure to the case of $k=4$. We fix ${\cal 
B}$, by picking one of the first 4 non-trivial entries in table 
\ref{NScharges} (as we want something that is not in our tables yet, we 
require ${\cal B}( \{p_i \} )\neq 0$). For $p=5$ the expression for ${\cal C}( 
\{p_i \} )$ is simple: equation (\ref{cform4}) has only one coefficient. To 
look for new configurations we have to set ${\cal C}( \{p_i \} ) =1$. 

Then we have the following cases:
\begin{itemize}
\item $b_1 =1$: Setting $c=1$ breaks supersymmetry.
\item $b_{12} =1$: Setting $c=1$ gives the model (0,0,12,4,2).
\item $b_{12} = b_3 = 1$: $c=1$ breaks supersymmetry.
\item $b_{12} = b_{34} =1$: $c=1$ breaks supersymmetry.
\end{itemize}

Hence in 6 dimensions, after a few straightforward computations we found one 
new model. This model made a brief appearance in \cite{deBoer}, and was more 
thoroughly described in \cite{Keurentjes01}. In the next subsection we will 
find models that have not appeared in the literature thus far. 

\subsection{4+1 dimensional theories}

Again we start by fixing ${\cal B}( \{p_i \} )$, by picking an entry from 
table \ref{NScharges}. The expression for ${\cal C}( \{p_i \} )$ is still 
relatively simple (see eq. \ref{cform5}). 

We will use a number of symmetries and properties of the expressions 
for ${\cal B}( \{p_i \} )$ to fix the polynomial ${\cal C}( \{p_i \} )$ to a 
standardform without affecting ${\cal B}( \{p_i \} )$. These coordinate 
transformations play an even more prominent role in fixing a standard form for 
the 4-dimensional theories.

A very useful and important property is, that the set of indices on the 
non-zero $b_{ij}$, $b_i$ has an upper bound. For $k=5$ this leads immediately 
to the following simplification. Let $m$ be the smallest integer 
that does not occur in the indices appearing on the non-zero entries of 
$b_{ij}$, $b_i$. If any of the $c_i$ with $i \geq m$ is non-zero, one first 
relabels coordinates (only coordinates with $i \geq m$ should be relabelled !) 
to set $c_m=1$. Subsequently one can transform  away all  $c_i$ with $i \neq 
m$, as well as $c$, by setting $x_{m} = {\cal C}(\{x_i \} )$, while leaving 
all other coordinates invariant. Because $c_{m}$ was non-zero, this 
transformation is non-singular, and because the coordinates $x_i$ with $i \neq 
m$ are not affected, this does not change the expression for ${\cal B}( \{p_i 
\} )$. We hence end up with a configuration given by ${\cal B}( \{p_i \} )$ 
and ${\cal C}( \{p_i \} ) = p_{m}$. Now applying a T-duality in the 
$m$-direction, we see that this configuration is related to one that already 
appeared one dimension higher. Summarizing: If any of the $c_i$ with $i \geq 
m$ is non-zero, the configuration is related to one considered before and 
therefore it is either ruled out, or already in our table. Therefore we can 
restrict to $c_i=0$ for $i \geq m$.

We now turn to the $c_i$, with $i < m$. If $b_{i,i+1}=1$ for some (odd) $i$, 
then one can assume that $c_{i+1} = 0$, because if it is not, we can set it 
to zero in the following way: If $c_i =0$ , $x_i 
\leftrightarrow x_{i+1}$ (other coordinates invariant) leaves ${\cal B}( 
\{p_i \} )$ invariant, but sets $c_{i+1}$ to zero; if $c_i=1$, setting 
$x_i \rightarrow x_i + x_{i+1} +1$ leaves ${\cal B}( \{p_i \} )$ 
invariant but removes $p_{i+1}$ dependence of ${\cal C}( \{p_i \} )$.

The previous trick will already eliminate many $c_i$'s, but we can still do 
more if we have $b_{12} = b_{34} =1$. In the way just described one sets $c_2$ 
and $c_4$ to zero, but one can also eliminate a non-zero $c_3$. Either 
$c_1=0$, and one uses $(x_1,x_2) \leftrightarrow (x_3,x_4)$; or $c_1 =1$ in 
case one uses $x_1 \rightarrow x_1 + x_3$, $x_4 \rightarrow x_2 + x_4$. In 
either case ${\cal B}( \{p_i \} )$ remains invariant, but $c_3$ is eliminated. 
Hence in this case we only have to consider non-zero $c_1$ and possibly $c_5$.

The following argument also presents a simplification in 5-dimension, but will 
in particular be a crucial shortcut to some computations in 4-dimensions. As 
remarked before, in models where the number of planes with NS charge is equal 
to the number of planes without NS charge, it is impossible to have 
$\widetilde{\textrm{O}p}{}^-$ planes while preserving supersymmetry. So, for 
these theories, instead of computing $r$ case by case, we can also simply 
demand absence of $\widetilde{\textrm{O}p}{}^-$ planes in these models. This 
amounts to requiring that, whenever ${\cal B}( \{p_i \} )$ gives the value $0$ 
at a certain plane, we should also require ${\cal C}( \{p_i \} )$ to give $0$ 
at that particular plane. Solving for this constraint fixes most, sometimes 
even all coefficients in the polynomial ${\cal C}( \{p_i \} )$.  

After these preliminary considerations we simply check the remaining options 
case by case: 

\begin{itemize} 
\item $b_1 =1$: As explained, we set $c_i =0$ for $i >1$.  Supersymmetry 
requires absence of $\widetilde{\textrm{O}4}{}^-$ planes. Then one quickly 
deduces that $c=0$ (because that would turn planes at $p_i=0$ into 
$\widetilde{\textrm{O}4}{}^-$). Hence the only non-trivial possibility is 
$c_1 = 1$, which gives $(16,0,0,16,0)$. \item $b_{12} =1$: We can set $c_i =0$ 
for $i >1$. then the only options left are:
\begin{enumerate} 
\item $c = 1$: Breaks supersymmetry. 
\item $c_1 = 1$: Leads to the model (16,0,8,8,4). 
\item $c_1 = c =1$: Leads to (8,8,16,0,0).
\end{enumerate}
\item $b_{12} = b_3 = 1$: Set $c_i = 0$ for $i >3$ and $c_2=0$. 
Supersymmetry requires absence of $\widetilde{\textrm{O}4}{}^-$ planes. Then 
one quickly deduces that $c=0$ (because that would turn planes at $p_1=p_3 =0$
into $\widetilde{\textrm{O}4}{}^-$) that $c_1 = 0$ (with planes at $p_1 = 1$, 
$p_3 =0$), and $c_3 =0$ (with planes at $p_1=p_3=1$). Hence supersymmetry 
requires ${\cal C}( \{p_i \} ) \equiv 0$, and we find no new models. 
\item $b_{12} = b_{34} =1$: Set $c_i =0$ for $i >
1$.
\begin{enumerate}
\item $c=1$: Breaks supersymmetry. 
\item $c_1 = 1$: Leads to (12,4,8,8,0). 
\item $c_1 = c = 1$: Breaks supersymmetry.
\end{enumerate}  
\item $b_{12} = b_{34} = b_5 =1$: Set $c_2 = c_3 =c_4 =0$.  Supersymmetry 
requires absence of $\widetilde{\textrm{O}4}{}^-$ planes. Then one quickly 
deduces that $c=0$ (because that would turn the plane at $p_i =0$ into 
$\widetilde{\textrm{O}4}{}^-$) and that $c_1 = 0$ (with planes at $p_1 = 1$), 
and $c_5 =0$ (with planes at $p_1 = p_5= 1$). Hence supersymmetry requires 
${\cal C}( \{p_i \} )\equiv 0$ and we find no new models.  \end{itemize}

With a little more effort than in the previous subsection, we have identified 
in total 4 new models. 

\subsection{3+1-dimensional theories}

In this dimension computations get very tedious. This is to some extent 
clear from the expression for ${\cal B}( \{p_i \} )$ (\ref{bform}) and ${\cal 
C}( \{p_i \} )$ (\ref{cform6}). In a very literal sense, they are both equally 
complicated, but after having fixed a standardform of ${\cal B}( \{p_i \} )$ 
there are only relatively few possibilities left to manipulate ${\cal 
C}( \{p_i \} )$.

One category of theories is still (relatively) simple to tackle: the ones with 
equal number of planes with NS charge and planes without NS charge. 
Supersymmetry imposes the absence of $\widetilde{Op^-}$ planes, and having 
realized that, the computation to be done is straightforward: 

\begin{itemize}
\item $b_1 =1$.

The absence of $\widetilde{\textrm{O}3}{}^-$ planes restricts the 
polynomial ${\cal C}( \{p_i \} )$ to 
\be
{\cal C}( \{p_i \} ) = p_1 (\sum_{j \neq 1} c_{1j} p_j + c_1)
\ee
If at least one of the $c_{1j}$ is non-zero, we use (if necessary) $x_2
\leftrightarrow x_j$ to set $c_{12}=1$. Subsequently we transform $x_2
\rightarrow (\sum_{j \neq 1} c_{1j} x_j + c_1)$. Hence we are left with the
following inequivalent possibilities:
\begin{enumerate}
\item $c_{12} = 1$: One can use T-duality in 2-direction to relate this to
$b_1=c_1=1$ for $p=4$
\item $c_1 = 1$: This leads to (32,0,0,32,0).
\end{enumerate}

\item $b_{12} = b_3 = 1$.

Absence of $\widetilde{\textrm{O}3}{}^-$ restricts the 
polynomial ${\cal C}( \{p_i \} )$ to 
\be
{\cal C}( \{p_i \} ) = c_{12} p_1 p_2 + c_{13} p_1 p_3 + c_{23} p_2 p_3 + c_3 p_3
\ee
with
\bd
c_{12} + c_{13}  + c_{23}  + c_3 = 0
\ed
This gives us 8 remaining possibilities to be checked. However, we still have 
some remaining symmetries that are generated by the transformations: $x_1 
\rightarrow  x_1 + x_2 +1$, other coordinates invariant; $x_ 1 \leftrightarrow 
x_2$, other coordinates invariant; and, $x_1 \rightarrow x_1 + x_2$, $x_3 
\rightarrow x_2 + x_3$. These leave ${\cal B}( \{p_i \} )$ invariant but have 
a non-trivial effect on ${\cal C}( \{p_i \} )$. Using these symmetries, one 
can reduce to 2 inequivalent options: 
\begin{enumerate} 
\item $c_{12} = c_{13} = 1$: Leads to (32,16,0,16,0). 
\item $c_{12} = c_3 =1$: Leads to (32,0,0,32,0) 
\end{enumerate}

\item $b_{12} = b_{34} = b_5 =1$.

Again we require absence of $\widetilde{\textrm{O}3}{}^-$ planes. A somewhat 
lengthy computation shows that this requirement restricts the polynomial 
${\cal C}( \{p_i \} )$ to \be {\cal C}( \{p_i \} ) = c_{12} p_1 p_2 + c_{34} 
p_3 p_4 + c_5 p_5 \ee with
\bd
c_{12} = c_{34} = c_5
\ed
Hence the only new model comes from setting $c_{12} = c_{34} = c_5 = 1$, which
gives (32,0,0,32,0).
\end{itemize}

The most tedious computations are the ones that will follow now. We found it 
convenient to invoke a simple computerprogram for parts of the analysis. 
Nevertheless, a number of computations was still done by hand. Equivalence or 
inequivalence of certain models can often be fairly easily checked, and allows 
us to avoid wasting computer time on configurations that have already been 
analyzed.

\begin{itemize}
\item $b_{12} =1$.
We use coordinate transformations in $x_i$ 
with $i=3,4,5,6$ to bring the part of ${\cal C}( \{p_i \} )$ with terms involving 
(only) these coordinates to standardform. 

We will start with the easiest case: 
\begin{enumerate} 
\item $c_{34} = c_{56} =1$.

Redefining $x_i$ ($i=3,4,5,6$) one can set $c_{1i}$, $c_{2i}$ and $c_i$ to 
zero. Using $x_1 \leftrightarrow x_2$, $x_1 \rightarrow x_1 + x_2 + 1$ also 
$c_2$ can be set to zero. The only possibly non-zero coefficients of ${\cal 
C}$ (apart from $c_{34}$ and $c_{56}$) are then $c_{12}$, $c_1$, $c$. This 
gives in total 8 models, but a computation reveals that all have $r <0 $, and 
therefore they are not supersymmetric.  

\item $c_{34} = 1$, $c_{56}=0$. 

We may set coefficients $c_{3i}$, $c_{4i}$, $c_3$ and $c_4$ to zero by 
redefining $x_3$ and $x_4$. 

If $c_{15} = c_{16} = c_{25} = c_{26} = 0$, one sets $c_2=0$ by the 
transformations $x_1 \leftrightarrow x_2$ and $x_1 \rightarrow x_1 + x_2 +1$. 
One can now also use $x_5 \leftrightarrow x_6$, and $x_5 \rightarrow x_5 + 
x_6$. to set $c_6$ to zero. If $c_5$ non-zero, one can transform to 
$c_1 = c=0$. Summarizing, with $c_{34} =1$, all $c_{ij}=0$ except possibly 
$c_{12}$, we only need to consider non-zero values for $c_{12}$, $c_1$, $c_5$ 
and $c$. Computing $r$ for these various possibilities one finds that $r \geq 
0$ for:
\begin{itemize} 
\item $c = 0$: This is related by T-duality in the 3 and 4 directions 
to a model considered previously.
\item $c_{12} = 1$: Leads to the model (36,4,12,12,2).
\end{itemize}

If $c_{15}=1$, one may set $c_{12}, c_{16}$ and $c_1$ to zero by
redefining $x_5$. With $c_{16}= 1$ but $c_{15} =0$ one uses $x_5
\leftrightarrow x_6$ .Thus in case of $c_{15} = c_{34}= 1$ we only need to 
consider non-zero $c_2$, $c_5$, $c_6$ and $c$. For these coefficients there is 
only one supersymmetric model:
\begin{itemize} 
\item $c_{15} = 1$: Leads to (32,8,16,8,0). 
\end{itemize}

With $c_{25}=1$ , but $c_{16}= c_{15}=0$, one uses $x_1 \leftrightarrow x_2$. 
With $c_{15}=c_{25}=1$, but $c_{16}=0$ one uses $x_1 \rightarrow x_1 + x_2 
+1$. When $c_{25}=c_{16} =1$ one can absorb any $c_{1i}$ ($i \neq 6$) and 
$c_1$ upon redefining $x_6$, and $c_{2i}$ ($i \neq 5$)and $c_2$ upon 
redefining $x_5$. Hence for this subclass of theories one only has to consider 
non-zero-values for $c_5$, $c_6$ and $c$. All of these turn out to result in 
$r < 0$. 

With $c_{26}=1$ but $c_{15}= c_{16}= c_{25}=0$ one uses $x_1 \leftrightarrow 
x_2$ to reduce to a previously considered case. If $c_{26}=c_{15}=1$, one uses 
$x_1 \leftrightarrow x_2$. If $c_{26}=c_{16}=1$ use $x_5 \leftrightarrow x_6$. 
If $c_{26} = c_{15} = c_{16} =1$ use $x_1 \leftrightarrow x_2$. 

In case $c_{26}=c_{25}=1$, redefine $x_5$ to absorb $c_{26}$. Any remaining 
model with $b_{12}=c_{34}=1$ can be transformed to one of the previously 
considered models, and we can pass on to the next class.

\item $c_{34} = c_{56} = 0$.

As long as $c_{ij}=0$, ${\cal C}( \{p_i \} )$ takes the simpler form (\ref{cform5}),
we can use the same set of tricks as in the $p=4$ case, and hence we only 
consider non-zero values for $c_1$ and $c$. This results in one supersymmetric 
model: 
\begin{itemize}
\item $c_1=1$: Leads to the model (32,0,16,16,0).
\end{itemize}

If $c_{12}=1$, we can set $c_2 =0$ by using $x_1 \leftrightarrow x_2$ and $x_1 
\rightarrow x_1 + x_2 + 1$. If $c_{12}= c_3 = 1$, one can absorb other $c_i$ 
and $c$ by redefining $x_3$. Considering the remaining possibilities gives the 
supersymmetric models:
\begin{itemize}
\item $c_{12}=1$: Leads to (48,0,0,16,8).
\item $c_{12}=c_1=1$: Leads to (32,16,16,0,0).
\end{itemize}

If at least one of the $c_{1j}$ ($j \neq 2$) is non-zero, one can use 
coordinate transformations in $x_i$ ($i = 3,...6$) to set all $c_{1j}$ except 
$c_{13}$, and $c_1$ to zero. If $c_2$ non-zero can set $c=0$ by redefining 
$x_2 \rightarrow x_1 + x_2 + 1$. Possible values for $c_i$ with $i > 3$ 
can all be absorbed in $c_4$ by using coordinate transformations in $x_i$ 
with $i=4,5,6$. Hence we capture all of these theories by setting $c_{13}=1$, 
and considering all values for $c_2$, $c_3$, $c_4$ and $c$. This leads to the 
following set of supersymmetric models.
\begin{itemize} 
\item $c_{13}=1$: Related to a higher dimensional model by using T-duality in 
the 3-direction. 
\item $c_{13}=c_3=1$: Also related by T-duality in the 3-direction to a higher 
dimensional model.
\item $c_{13}=c_3 = c_2 =1$: Leads to (32,0,16,16,0).
\end{itemize}

We next consider $c_{23}=1$, $c_{13}=0$; in this case one uses $x_1 
\leftrightarrow x_2$. If $c_{23}=c_{13}=1$, use $x_1 
\rightarrow x_1 + x_2 +1$. 

If $c_{23}= c_{14}=1$, one can redefine $x_3$ and $x_4$ to absorb all other
$c_{1j}$, $c_{2j}$, $c_1$ and $c_2$. If either one of $c_3$ or $c_4$ is zero, 
they can both be set to zero in the following way: If $c_3=0$ and $c_4=1$, use 
$x_1 \rightarrow x_1 + x_2 +1, x_3 \rightarrow x_3 + x_4$; If $c_3=1$ and 
$c_4=0$, use $x_2 \rightarrow x_1 + x_2 +1, x_4 \rightarrow x_3 + x_4$. Hence 
in this case we only need to check for values of $c_3 =c_4$ and $c$. This 
results in a single supersymmetric model  
\begin{itemize} 
\item $c_{23} = c_{14} =1$: Leads to (32,8,16,8,0) 
\end{itemize}

If $c_{23} =1$ and $c_{15}=1$ and/or $c_{16}=1$ one uses coordinate 
transformations in $x_4, x_5, x_6$ to set $c_{14}=1$ and proceeds as 
previously.
  
If $c_{24}=1$ and $c_{23}=0$ one uses $x_3 \leftrightarrow x_4$. 
If $c_{24}=c_{23} =1$ then redefine $x_3$ to absorb all other $c_{2j}$ and 
$c_2$, and hence set $c_{24}=0$. Finally, if $c_{25}$ or $c_{26}$ is non-zero, 
one again permutes $x_4,x_5,x_6$, such that $c_{24}$ is set to one, and 
proceeds as before.
\end{enumerate} 

This finishes the computation of configurations with $b_{12}=1$

\item $b_{12} = b_{34} =1$.

We would like to bring the part with quadratic terms of the expression for 
${\cal C}(\{p_i \})$ to standardform, but our freedom is limited to the 
coordinates $x_5$ and $x_6$. Therefore the ``standardform'' simply amounts to 
stating the value of $c_{56}$, all other symmetries have to come from the 
transformations that leave ${\cal B}(\{ p_i\})$ invariant.

\begin{enumerate} 
\item $c_{56} =1$.

In principle one could go through the whole tedious procedure of the fixing of 
symmetries. However, running a computer program computing all possible 
configurations with $b_{12} = b_{34}= c_{56}=1$ (without actually fixing any 
other symmetries) shows that none of these theories is supersymmetric. 
Therefore our efforts would be in vain anyway, and we simply pass on to 
$c_{56}=0$. 

\item $c_{56} =0$. 

At first we set $c_{i5}, c_{j6}=0$. 

Symmetries of the B-field configuration are respected by a group of symmetries 
(see $p=4$), generated by
\bea
(x_1,x_2,x_3,x_4) & \rightarrow & (x_2,x_1,x_3, x_4) \\
(x_1,x_2,x_3,x_4) & \rightarrow & (x_1 +x_2 +1,x_2,x_3, x_4) \\
(x_1,x_2,x_3,x_4) & \rightarrow & (x_3,x_4,x_1, x_2) \\
(x_1,x_2,x_3,x_4) & \rightarrow & (x_1 + x_3, x_2, x_3, x_2 + x_4) 
\eea
Of course there are also symmetries involving $x_5$ and $x_6$, but because 
${\cal B}(\{ p_i \})$ does not involve these coordinates, we do not mention 
them at this point.

The set of coefficients $c_{12}, c_{13}, c_{14}, c_{23}, c_{24}$ and $c_{23}$ 
can a priori take 64 values, but under the above symmetries many of these are 
related. We decompose the orbits of the symmetries on the space of possible 
values of the set of above $c_{ij}$'s (there turn out to be 6 of these) and 
pick a representative from each orbit. Then we run a computer calculation on 
each representative, checking on supersymmetric configurations if combinations 
of $c_i$'s and $c$ are turned on. This immediately leads to the following 
results: 

\begin{itemize} 
\item $c_{ij}=0$: If not $c_i=c=0$ then supersymmetry is broken. The remaining 
configuration has ${\cal C}(\{ p_i\})=0$ and can be found in section \ref{B}.
\item $c_{12}=1$: Only $c_i=c=0$ is a supersymmetric configuration; this leads 
to (36,12,4,12,2). 
\item $c_{13}=1$: Supersymmetric configurations are $c_i=0$, $c_1=1$ 
and $c_3=1$. These are equivalent by symmetries: $x_1 \rightarrow x_1 + x_3, 
x_4 \rightarrow x_2 + x_4$, takes $c_{13}=1, c_{i}=0$ to $c_{13}=c_3=1$, while 
$x_3 \rightarrow x_1 + x_3, x_2 \rightarrow x_2 + x_4$, takes $c_{13}=1, 
c_{i}=0$ to $c_{13}=c_1=1$. Therefore they are really only one model, with 
characteristic (32,16,8,8,0).
\item $c_{13}=c_{24}=1$: Supersymmetric configurations are $c_i=0$, 
$c_2=c_3=1$ and $c_1=c_4=1$. These are also related to each other by 
the symmetries $x_1 \rightarrow x_1 + x_3, x_4 \rightarrow x_2 + x_4$, and 
$x_3 \rightarrow x_1 + x_3, x_2 \rightarrow x_2 + x_4$, and hence all lead 
to the same model, (32,8,8,16,0). 
\item $c_{12}=c_{13}=c_{24}=1$: None of the theories in this orbit is 
supersymmetric.
\item $c_{12}=c_{34}=1$: Only $c_1=c=0$ is supersymmetric. This leads to the 
model (40,0,0,24,4). 
\end{itemize}

Next consider $c_{15}=1$.  We can now absorb other $c_{1i}$ and $c_1$ by 
redefining $x_5$. Combining this with the orbit structure one is left with: 
\begin{itemize} 
\item $c_{15}=1$: The only supersymmmetric model has $c_i=c=0$, but 
is related by T-duality to a higher dimensional model. 
\item $c_{15}=c_{24}=1$: All values for $c_i$ and $c$ give tadpoles. 
\item $c_{15}=c_{34}=1$: The only supersymmetric model is $c_i=c=0$, leading 
to (32,8,8,16,0). 
\end{itemize} 

If $c_{16}=1$ we use $x_5 \leftrightarrow x_6$. If $c_{15}=c_{16}=1$ one 
absorbs $c_{16}$ by redefining $x_5$. 

Next we turn to $c_{25}=1$. Now we absorb all other $c_{2i}$ and 
$c_2$ in a redefinition of $x_5$
\begin{itemize} 
\item $c_{25}=1$: use $x_1 \leftrightarrow x_2$. 
\item $c_{13}=c_{25}=1$: use $x_1 \leftrightarrow x_2, x_3 \leftrightarrow x_4$. 
\item $c_{34}=c_{25}=1$: use $x_1 \leftrightarrow x_2$. 
\end{itemize} 
If $c_{15}=c_{25}=1$, one first absorbs the other $c_{1i}$ and $c_{2i}$ in a
redefinition of $x_5$. Then use $x_1 \rightarrow x_1 + x_2 +1$ to eliminate 
$c_{25}$. 

If $c_{25}=c_{16}=1$, one first absorbs the other $c_{1i}$ and $c_{2j}$ in 
redefinitions of $x_5$ and $x_6$. The only two-index coefficient in ${\cal 
C}(\{p_i\})$ that is not fixed yet is $c_{34}$, but a check reveals that both 
$c_{34}=0$ and $c_{34}=1$ always result in supersymmetry breaking 
configurations. When $c_{25}=c_{16}=c_{15}=1$ one again absorbs $c_{1i}$ and 
$c_{2i}$ (with $i \neq 5$) in a redefinition of $x_5$, and proceeds as 
previously.  With $c_{26}=1,c_{25}=0$, use $x_5 \leftrightarrow x_6$. If 
$c_{26}=c_{25}=1$ : Redefine $x_5$ to absorb $c_{26}$.

Next consider $c_{35}=1$. Absorb all other $c_{3i}$ in a redefinition of 
$x_5$. The following options remain:
\begin{itemize}
\item $c_{35}=1$: Use $x_1 \leftrightarrow x_3$, $x_2 \rightarrow x_4$.
\item $c_{12}=c_{35}=1$: Use $x_1 \leftrightarrow x_3, x_2 \leftrightarrow 
x_4$.
\item $c_{24}=c_{35}=1$: Use $x_1 \leftrightarrow x_3, x_2 \leftrightarrow 
x_4$. 
\item $c_{12}=c_{24}=c_{35}=1$: These all break supersymmetry.
\end{itemize}
If $c_{35}=c_{15}=1$, one absorbs all other $c_{3i}$ and $c_{1i}$ in a 
redefinition of $x_5$. The only coefficient left to consider is $c_{24}=0,1$, 
but now one can always use $x_1 \rightarrow x_1 + x_3, x_4 \rightarrow x_2 + 
x_4$ to relate this to a previously considered model. With $c_{35}=c_{16}=1$. 
Again one absorbs all other $c_{1i}$ and $c_{3i}$ in redefinitions of $x_5$ 
and $x_6$, and one is left with only $c_{24}=0,1$. All possible models with 
these coefficients break supersymmetry.

With $c_{35} = c_{25}=1$ absorb all other $c_{2i}$ and $c_{3i}$ in a 
redefinition of $x_5$. Then use $x_2 \leftrightarrow x_2+ x_3, x_4 \rightarrow 
x_1 + x_4$. The cases $c_{35}=c_{25}=1$ with $c_{15}, c_{16}=0,1$ are treated 
similarly. With $c_{35}=c_{26}=1$, first transform away $c_{2i}$ and $c_{3i}$, 
then use $x_1 \leftrightarrow x_2$. With $c_{35}=c_{26}=c_{25}=1$, transform 
away $c_{2i}, c_{3i}$ and $c_{26}$. With $c_{36}=1$ use $x_5 \leftrightarrow 
x_6$. With $c_{35}=c_{36}=1$, absorb $c_{36}$ in a redefinition of $x_5$.

With $c_{45}=1$ absorb all other $c_{4i}$ in a redefinition of $x_5$. After 
this one is left with possible non-zero values for $c_{12}$ and $c_{13}$. By 
using $x_1 \leftrightarrow x_4, x_2 \leftrightarrow x_3$ this can always be 
reduced to a previously considered model. With $c_{45}=c_{15}=1$ one redefines 
$x_5$ to absorb other $c_{1i}$ and $c_{4i}$. Then use $x_1 \rightarrow x_1 + 
x_4, x_3 \rightarrow x_2 + x_3$. With $c_{45}=c_{16}=1$, absorb other $c_{1i}$ 
and $c_{4i}$ in redefinitions of $x_5$ and $x_6$. Then use $x_1 
\leftrightarrow x_2, x_3 \leftrightarrow x_4$. With $c_{45}=c_{25}=1$, first 
absorb $c_{2i}$ and $c_{4i}$ in a redefinition of $x_5$. Then use $x_2 
\rightarrow x_2 + x_4, x_3 \rightarrow x_1 + x_3$. With $c_{45}=c_{25}=1$ and 
$c_{15}$ and/or $c_{16}$ is $1$, absorb all other $c_{1i}$, $c_{2i}$ and 
$c_{4i}$ with coordinate definitions of $x_5$ and $x_6$, and use again $x_2 
\rightarrow x_2 + x_4, x_3 \rightarrow x_1 + x_3$. With $c_{45}=c_{26}=1$, use 
$x_1 \leftrightarrow x_2$, $x_3 \leftrightarrow x_4$.

With $c_{45}=c_{35}=1$, absorb all other $c_{3i}$ and $c_{4i}$ in a 
redefinition of $x_5$, and then use $x_3 \rightarrow x_3 + x_4 + 1$. With 
$c_{45}=c_{36}=1$, absorb all other $c_{3i}$ and $c_{4i}$ and use $x_1 
\leftrightarrow x_3, x_2 \leftrightarrow x_4$. The same for 
$c_{45}=c_{36}=c_{35}=1$.

Finally, for $c_{46}=1$ and $c_{45}=0$, one uses $x_5 \leftrightarrow x_6$, 
and for $c_{46}=c_{45}=1$ one can use $x_5 \rightarrow x_5 + x_6$.
\end{enumerate}

We have explicitly or implicitly considered all models with $b_{12}=b_{34}=1$, 
and we pass on to the last category of theories.
  
\item $b_{12} = b_{34} = b_{56} = 1$. 

There are only relatively few, to be precise 4 D3 branes to be 
possibly distributed over the O3 planes. One possibility is requiring absence 
of $\widetilde{\textrm{O}3}{}^-$ planes, which automatically implies 
$c_{12}=c_{34}=c_{56}$, all others zero. Therefore, the only non-trivial 
solution to this constraint is  
\begin{itemize} 
\item $c_{12}=c_{34}=c_{56}=1$ leading to (36,0,0,28,2) 
\end{itemize} 
Running a check with the computer on other configurations with $b_{12} = 
b_{34} = b_{56} = 1$ immediately reveals that there are no other 
supersymmetric solutions, than the above one and the one with ${\cal 
C}(\{p_i\})=0$. 
\end{itemize}

In this subsection we have identified in total 17 new models, that were 
collected in table \ref{mixedb}.

\section{S-duality in 4 dimensions} \label{S}

All our orientifold theories on $T^6/\Z_2$ result in low energy effective 4 
dimensional \mbox{${\cal N}=4$} supersymmetric Yang-Mills theories. As is well 
known, these theories exhibit an $SL(2,\Z)$ S-duality symmetry. This provides 
an interesting check on our results and methods.

First of all we note that the S-duality symmetry was already manifest in our 
formalism. In 4 dimensions the RR charges are described by (\ref{cform6}), 
which is isomorphic to the formula for the NS charges (\ref{bform}). The group 
$SL(2,\Z)$ is generated by
\be \label{sgen}
S= \left( \ba{rr} 0 & -1 \\ 1 & 0 \ea \right) \qquad T = \left( \ba{rr} 
1 & \phantom{-}1 \\ 0 & 1 \ea \right)
\ee
Because the formula's (\ref{bform}) and (\ref{cform6}) are over the field 
$\Z_2$, the manifest S-duality group is not $SL(2,\Z)$, but its reduction 
modulo 2, $SL(2,\Z_2)$. This group is generated by matrices $S$ and $T$, 
reduced modulo 2. The effect of the matrix $S$ is the interchange of the 
formula's (\ref{bform}) and (\ref{cform6}). The transformation $T$ leaves 
(\ref{cform6}) invariant, but replaces (\ref{bform}) by \be \label{bcform}
({\cal B}+{\cal C})(\{p_i \})  \equiv \int_{M_p} (B +C) = 
\sum_{i < j} (b_{ij}+c_{ij}) p_i p_j + \sum_i (b_i+c_i) p_i + (b +c) 
\ee

Recall that ${\cal B}(\{p_i \})$ gave 1 on O$3^+$ 
and $\widetilde{\textrm{O}3}{}^+$, and 0 on O$3^-$ and 
$\widetilde{\textrm{O}3}^-$, whereas ${\cal C}(\{ p_i \})$  resulted in 1 on  
$\widetilde{\textrm{O}3}^+$ and $\widetilde{\textrm{O}3}{}^-$, and 0 on 
O$3^+$ and  O$3^-$. Hence formula (\ref{bcform}) gives the value 1 on O$3^+$ 
and $\widetilde{\textrm{O}3}{}^-$, and 0 on O$3^-$ and 
$\widetilde{\textrm{O}3}{}^+$.

From the action of $S$ and $T$ on the formula's describing the charges we see 
that $S$ interchanges O$3^+$ with 
$\widetilde{\textrm{O}3}{}^-$, and therefore $n_+$ with 
$\tilde{n}_-$, while $T$ interchanges O$3^+$ with 
$\widetilde{\textrm{O}3}{}^+$, and hence $n_+ \leftrightarrow \tilde{n}_+$. 
These two transformations generate the whole permutation group ${\cal S}_3 
\cong SL(2,\Z_2)$, acting on the 3 numbers in the triple $(\tilde{n}_-, n_+, 
\tilde{n}_+)$.

One can now distinguish essentially 3 possibilities. First, if $\tilde{n}_- = 
n_+ = \tilde{n}_+$, then the configuration forms a singlet under $SL(2,\Z_2)$. 
There are actually 2 models in our tables with this property: the trivial 
compactification, with (64,0,0,0,64), and a much more interesting model, with 
(40,8,8,8,4).

Instead of the 3 numbers $\tilde{n}_-$, $n_+$, $\tilde{n}_+$ being all equal, 
they could also be all different. This possibility is however not realized at 
all, as can be seen by inspection of our tables. We don't know whether this is 
just a numerical coincidence, or there is a deeper level of understanding of 
this fact possible. 

All configurations in our tables that are not singlets under $SL(2,\Z_2)$ have 
2 out of the 3 numbers $\tilde{n}_-$, $n_+$, $\tilde{n}_+$ equal. In this 
case, some transformations in $SL(2,\Z_2)$ have a trivial effect on the triple 
$(\tilde{n}_-, n_+, \tilde{n}_+)$. Only a $\Z_3$ subgroup is manifest, and the 
orbits of the group consist of 3 elements.

It is a nice consistency check on our results to divide the models in 
our tables in orbits under $SL(2,\Z_2)$. Closure of all orbits gives us added 
confidence that our classification is indeed complete. In most cases this is a 
fairly simple exercise, as often the set of numbers $(\tilde{n}_-, n_+, 
\tilde{n}_+)$ specifies the model completely (in 4 dimensions $n_- = 64
-\tilde{n}_- - n_+ - \tilde{n}_+$, and $r$ follows from (\ref{tad})). 

There exist however models with equal $(n_-, n_+, \tilde{n}_-, \tilde{n}_+, r)$ 
that are nevertheless not equivalent. Earlier examples of models with this 
property involved O$p^-$ and O$p^+$ planes only \cite{Bergman, deBoer, 
Keurentjes01} (see our table \ref{NScharges}), but we now see that this 
phenomenon also occurs in the mixed cases.  We now also need the information 
provided by the coefficients of ${\cal B}(\{p_i\})$ and ${\cal C}(\{ p_i \})$ 
as these specify the geometry of all planes.

In this case we always have $r=0$ (we do not know whether there 
is a deeper reason that such models should have $r=0$, it follows from 
inspection of the tables). For some configurations whose $SL(2,\Z_2)$ orbit 
cannot be determined by the set of numbers $(n_-, n_+, \tilde{n}_-, 
\tilde{n}_+, r)$, we have attached an index to the number $r=0$ in tables 
\ref{mixeda} and \ref{mixedb}. Theories with the same index belong to the same 
$SL(2,\Z_2)$ orbit. The theories in table \ref{mixedb} with $r=0$, but not 
labelled by an index are S-dual to theories in tables \ref{NScharges} and 
\ref{RRcharges}. The reader should have no difficulty identifying these models 
and their duals.  

Rather interestingly, our whole formalism is $SL(2, \Z)$ dual, 
regardless whether in the end one finds a supersymmetric theory or not, 
and therefore also assigns dual theories to non-supersymmetric theories. 
It would be interesting to see to what extent physics in these 
non-supersymmetric ``dual'' theories is related to each other, the more 
because after the anti-brane annihilation process that turns 
$\widetilde{\textrm{O}p}{}^-$ into a bound state of O$p^-$ with an anti 
D$p$-brane, the manifest duality between various descriptions will have 
dissappeared.

\section{Conclusions} \label{sum}

By computing the holonomies of gerbes defined by the discrete charges of 
the orientifold planes, the problem of classification of orientifold 
planes on $T^k/\Z_2$ can be reduced to rather simple polynomial 
equations over the field $\Z_2$. Moreover, these polynomials represent a very 
compact way of storing information about the orientifold model, as many 
relevant quantities can be deduced from them. The techniques we used have 
their roots in \cite{deBoer}, where they were applied to K3's of the 
form $T^4/G$ ($G$ a finite group), but are useful in the present problem as 
well. In principle they may be relevant to any situation involving orientifold 
and/or orbifold fixed point sets with more than one connected component, in 
the presence of discrete torsion and similar degrees of freedom.

Preservation of supersymmetry reduces to a simple inequality. With these 
powerful techniques the classification of supersymmetric orientifolds was 
carried out for all $k \leq 6$.  The results can be found in the tables 
\ref{NScharges}, \ref{RRcharges}, \ref{mixeda} and \ref{mixedb}. 

All our theories are dual to toroidal compactifications of either type I theory, or 
the type IIB theory on $S^1/\delta \Omega$, with suitable holonomies on the 
torus. This has not played a large role in our analysis in this paper. Instead 
we refer the reader to \cite{deBoer, Keurentjes00a, Keurentjes01, Sen, 
Witten97} for various ideas and results.

For the benefit of the reader, we collect some information from our 4 tables 
in another table summarizing our results.

\TABLE{
\begin{tabular}{|c||ccccc||c|} \hline 
$d$ & $r=16$ & $r=8$ & $r=4$ & $r=2$ & $r=0$ & total \\
\hline
9 & 1 & 0 & 0 & 0 & 0 & 1 \\
8 & 1 & 0 & 0 & 0 & 1 & 2 \\
7 & 1 & 1 & 0 & 0 & 1 & 3 \\
6 & 1 & 1 & 0 & 0 & 2 & 4 \\
5 & 1 & 2 & 1 & 1 & 2 & 7 \\
4 & 1 & 2 & 2 & 1 & 7 & 13 \\
3 & 1 & 3 & 4 & 6 & 21 & 35 \\
\hline
\end{tabular}
\caption{Number of orientifold theories in $d$ dimensions for all values of 
$r$} \label{tottable} }

Table \ref{tottable} lists the number of maximally supersymmetric 
orientifold models in various dimensions, where $d$ is the number of spatial 
dimensions. It should be stressed that this is not a classification of 
irreducible components in the string moduli space. On the one hand, there are 
many components in the string moduli space that do not have an orientifold 
description. On the other hand, it is known that a single irreducible 
component in the string moduli space can give rise to multiple orientifold 
descriptions \cite{deBoer}. 

One of the things that immediately draws attention is the explosive growth of 
number of theories for $d < 6$. This can be viewed as the analogue of the 
existence of triples, and other non-standard compactifications in gauge 
theory, where for sufficiently high compact dimension there are many 
irreducible components in the moduli space of flat connections 
\cite{Borel, Kac, Keurentjes99}. Another typical fact is that theories 
with $r=4$ and $r=2$ appear simultaneously, at $d=5$, but that then 
subsequently there are no new values for $r$ encountered in the rest of 
the table. Furthermore, except for $r=0$, the numbers $r$ are even powers of 
2. Comparing with for example \cite{Borel, Keurentjes00a}, one sees that this 
does not follow from the possible bundles on $T^k$ (as there exist flat 
$Spin(32)/\Z_2$ bundles with rank 12 and rank 5, for example), and not even 
from supersymmetry (the previously mentioned bundles give perfectly valid 
vacua for supersymmetric gauge theories on tori), but from constraints that 
are intrinsic to string theory \cite{deBoer}. It would be nice to understand 
the numerology better\footnote{In \cite{Bergman} the rank 5 and rank 12 models 
are interpreted as string theories without supersymmetry. This 
involves the discrete cosmological constant of ref. \cite{Hyakutake}. In spite 
of the presence of such a cosmological constant the authors of \cite{Bergman} 
seem to assume a flat background away from the orientifold planes. It is not 
clear to this author that this is self-consistent.}. 

We also discussed the implications of S-duality of 4-dimensional 
${\cal N}=4$ supersymmetric gauge theories. Among the orientifold 
configurations we found, there are two singlets under the $SL(2,\Z)$ 
duality group, one with $r=16$ (64,0,0,0,16) and one with $r=4$ 
(40,8,8,8,4). Especially the latter one may be an interesting theory to 
study, the behavior under S-duality suggests a highly symmetric 
spectrum, and with $r=4$ gauge groups of relatively large rank are 
possible. Note that it is even possible to occupy at least one of each 
kind of O$p$-plane with a D$p$-brane pair. It may also be interesting to 
compute the lattice for the heterotic version of this theory, along the lines 
of \cite{deBoer}. The remaining theories organize in orbits of each 3 
elements, which is reflected in table \ref{tottable}, because if one omits 
the singlets, all the numbers in the last line of the table are divisible by 
3.

Although discussed in less detail, it should also be clear that the present 
techniques provide a rich source of examples of non-supersymmetric theories, 
similar to the ones analyzed in \cite{Schwarz, Sugimoto}.

To take this classification to still lower dimension, we need new 
insights concerning the rewriting of (\ref{cform7}) to some kind of 
standardform. An estimate on the needed time to perform a 
computer analysis without any new information, does not give much hope that 
continuing this classification to $p=2$ by brute force can be done in the 
near future.

\acknowledgments

I would like to thank Jan de Boer, Robbert Dijkgraaf, Amihay Hanany and 
Bernard Julia for helpful conversations, and Niels Keurentjes for valuable 
assistance with computer programming. This work is partly supported by EU 
contract HPRN-CT-2000-00122.

\end{document}